# Dimension Extractors and Optimal Decompression

David Doty*
Department of Computer Science
Iowa State University
Ames, IA 50011, USA
ddoty@iastate.edu

**Abstract**

A *dimension extractor* is an algorithm designed to increase the effective dimension – i.e., the amount of computational randomness – of an infinite binary sequence, in order to turn a "partially random" sequence into a "more random" sequence. Extractors are exhibited for various effective dimensions, including constructive, computable, space-bounded, time-bounded, and finite-state dimension. Using similar techniques, the Kučera-Gács theorem is examined from the perspective of decompression, by showing that every infinite sequence $S$ is Turing reducible to a Martin-Löf random sequence $R$ such that the asymptotic number of bits of $R$ needed to compute $n$ bits of $S$, divided by $n$, is precisely the constructive dimension of $S$, which is shown to be the optimal ratio of query bits to computed bits achievable with Turing reductions. The extractors and decompressors that are developed lead directly to new characterizations of some effective dimensions in terms of optimal decompression by Turing reductions.

## 1 Introduction

Effective dimension [33, 34] and strong dimension [2] are effectivizations of classical Hausdorff [21] and packing [49, 50] dimension, which can each be characterized in terms of betting strategies called *martingales*. By placing resource bounds on the martingales, individual infinite binary sequences can be assigned a non-zero effective dimension, interpreted as the *density of computational randomness* of the sequence. A *dimension extractor* is an algorithm designed to increase the effective dimension of a sequence. Like their counterparts in computational complexity theory [46], dimension extractors transform a source of

*This research was funded in part by grant number 9972653 from the National Science Foundation as part of their Integrative Graduate Education and Research Traineeship (IGERT) program.

weak randomness into a source of strong randomness, the difference being that the *algorithmic* randomness of a *sequence* is being extracted, rather than *classical* randomness of a *probabilistic source*.

To attack (but not settle) a question raised by Reimann [41], Terwijn, Miller, and Nies [38] concerning the ability of Turing reductions to increase the constructive dimension [34] of a sequence, we exhibit a constructive dimension extractor by showing that every sequence of positive constructive dimension is Turing equivalent to a sequence of constructive strong dimension [2] arbitrarily close to 1. The reduction is uniform with respect to the input sequence: a single oracle Turing machine, taking as input a rational upper bound on the dimension of the input sequence, works for every input sequence of positive constructive dimension. The construction of this extractor follows in a straightforward manner from an earlier result of Ryabko [42, 43] on constructive dimension and compression, and from Mayordomo's [36] characterization of constructive dimension in terms of Kolmogorov complexity (cf. [30]).

We then develop new techniques to extend and improve this result for effective dimensions at lower levels of computability; in particular, our first main result shows that computable, $\mathrm{p}_i$space, and $\mathrm{p}_i$ dimension [33] can be extracted using truth-table reductions, $\mathrm{p}_i$space Turing reductions, and $\mathrm{p}_i$ Turing reductions, respectively. The $\mathrm{p}_i$ hierarchy was defined by Lutz [31] as a hierarchy of classes of functions computable in super-polynomial, but sub-exponential, space or time. For instance, $\mathrm{p}_1$space is simply pspace, the class of functions computable in space $n^k$ for some constant $k$, and $\mathrm{p}_2$space is the class of functions computable in space $n^{(\log n)^k}$. It is also shown that finite-state dimension [2, 12] can be extracted with information lossless finite-state transducers [26]. Thus, with respect to constructive, computable, $\mathrm{p}_i$space, $\mathrm{p}_i$, and finite-state information density, any sequence in which *almost every* prefix has information density *bounded away from zero* can be used to compute a sequence in which *infinitely many* prefixes have information density that is *nearly maximal*. Furthermore, in the case of all dimensions except constructive dimension, in addition to the strong dimension extraction, a lower bound of $d/D - \epsilon$ is derived on the dimension of the extracted sequence, where $d$ and $D$ are the dimension and strong dimension, respectively, of the input sequence, and $\epsilon$ is an arbitrarily small positive constant. It follows that for any *regular* sequence (a sequence in which the dimension and strong dimension agree), both strong dimension *and* dimension are nearly optimally extracted.

For *all* dimensions, the intuition behind the proof is the same. The extractor acts as a compressor that compresses the input sequence close to its optimal compression ratio under the resource bound $\Delta$, which is precisely the $\Delta$-dimension of the sequence. It is well-known [30] that the shortest program to produce a finite string must itself be incompressible (i.e., have maximal Kolmogorov complexity). Mayordomo's Kolmogorov complexity characterization of constructive dimension [36], Hitchcock's $\Delta$-bounded Kolmogorov complexity characterization of $\Delta$-dimension for $\Delta$ = comp or $\mathrm{p}_i$space [22], López-Valdés and Mayordomo's $\mathrm{p}_i$ compressor/decompressor characterization of $\mathrm{p}_i$ dimension [51], and Dai, Lathrop, Lutz, and Mayordomo's compression characterization of finite-state dimension [12], are invoked to show that a compressed representation of a sequence must itself be more incompressible than the sequence and thus have higher dimension. Of course, this technique works for any dimension that has a characterization in terms of Kolmogorov complexity, such as, for instance, exponential time or exponential space bounded dimension.

In each case, the extractor is no more powerful than the resource bound defining the dimension, which is necessary to make the results non-trivial. For instance, without access to any oracle sequence, but given exponential space, a program can diagonalize against all pspace-bounded martingales to compute a pspace-random sequence. An extractor is interesting only when it has no more computational power than the class of algorithms it is trying to fool: all the randomness present in the output sequence must originate from the input sequence, and the extractor merely acts as a filter that distills the randomness out from the redundancy.

We then examine the Kučera-Gács theorem from the perspective of decompression. Kučera [28, 29] and Gács [18] independently showed that every infinite sequence is Turing reducible to a Martin-Löf random sequence. Our second main result extends this theorem by showing that every infinite sequence $S$ is Turing reducible to a Martin-Löf random sequence $R$ such that the asymptotic best-case and worst case number of bits of $R$ needed to compute $n$ bits of $S$, divided by $n$, are precisely the constructive dimension and constructive strong dimension, respectively, of $S$. We show that this is the optimal ratio of query bits to computed bits achievable with Turing reductions.

As an application of these results and techniques, the resource-bounded extractors are used to characterize the computable dimension of individual sequences in terms of decompression via truth-table reductions and to characterize the $\mathrm{p}_i$space dimension of individual sequences in terms of decompression via $\mathrm{p}_i$space-bounded Turing reductions, and the optimal decompression result is used to characterize constructive dimension in terms of decompression via Turing reductions.

The paper is organized as follows. Section 2 explains notation and background material. Section 3 introduces the concepts of compression and decompression via reductions, and also gives definitions and background for effective dimension. Section 4 exhibits dimension extractors for various effective dimensions. Section 5, which appeared in preliminary form as [13], uses techniques developed in section 4 to show that every sequence is optimally decompressible from a random one. Section 6 uses results of sections 4 and 5 to prove new characterizations of constructive, computable, and space-bounded dimensions.

# 2 Preliminaries

We refer the reader to [30] for an introduction to Kolmogorov complexity and algorithmic information theory, [48] for an introduction to computability theory, and [40] for an introduction to computational complexity theory.

## 2.1 Notation

All logarithms are base 2. We write $\mathbb{R}$, $\mathbb{Q}$, $\mathbb{Q}_2$, $\mathbb{Z}$, and $\mathbb{N}$ for the set of all reals, rationals, dyadic rationals, integers, and non-negative integers, respectively. For all $A \subseteq \mathbb{R}$, $A^+$ denotes $A \cap (0, \infty)$. $\{0,1\}^*$ denotes the set of all finite, binary *strings*. For all $x \in \{0,1\}^*$, $|x|$ denotes the *length* of $x$. $\lambda$ denotes the empty string. Let $s_0, s_1, s_2, \ldots \in \{0,1\}^*$ denote the standard enumeration of binary strings $s_0 = \lambda, s_1 = 0, s_2 = 1, s_3 = 00, \ldots$. For $k \in \mathbb{N}$, $\{0,1\}^k$ denotes the set of all strings $x \in \{0,1\}^*$ such that $|x| = k$. $\mathbf{C} = \{0,1\}^\infty$ denotes the *Cantor space*, the set of all infinite, binary *sequences*. For $x \in \{0,1\}^*$ and $y \in \{0,1\}^* \cup \mathbf{C}$, $xy$ denotes the concatenation of $x$ and $y$, $x \sqsubseteq y$ denotes that $x$ is a *prefix* of $y$; i.e., there exists $u \in \{0,1\}^* \cup \mathbf{C}$ such that $xu = y$, and $x \sqsubset y$ denotes that $x \sqsubseteq y$ and $x \neq y$. For $S \in \{0,1\}^* \cup \mathbf{C}$ and $i, j \in \mathbb{N}$, $S[i]$ denotes the $i^{\text{th}}$ bit of $S$, with $S[0]$ being the leftmost bit, $S[i \mathinner{..} j]$ denotes the substring consisting of the $i^{\text{th}}$ through $j^{\text{th}}$ bits of $S$ (inclusive), with $S[i \mathinner{..} j] = \lambda$ if $i > j$, and $S \upharpoonright i$ denotes $S[0 \mathinner{..} i-1]$. A *language* is a subset of $\{0,1\}^*$, and we identify a language $L \subseteq \{0,1\}^*$ with its *characteristic sequence* $\chi_L \in \mathbf{C}$, where the $n^{\text{th}}$ bit of $\chi_L$ is 1 if and only if $s_n \in L$, writing $L \upharpoonright i$ to denote $\chi_L \upharpoonright i$.

## 2.2 Kolmogorov Complexity and Coding

Fix a universal self-delimiting Turing machine $U$. Let $w \in \{0,1\}^*$. The *Kolmogorov complexity* of $w$ is

$$\mathrm{K}(w) = \min_{\pi \in \{0,1\}^*} \{\, |\pi| \mid U(\pi) = w \,\}.$$

The quantity $\frac{\mathrm{K}(w)}{|w|}$ is called the *Kolmogorov rate* of $w$. Given a time- and space-constructible bound $t : \mathbb{N} \to \mathbb{N}$, the *t-time-bounded Kolmogorov complexity* of $w$ is

$$\mathrm{K}^t(w) = \min_{\pi \in \{0,1\}^*} \{\, |\pi| \mid U(\pi) = w \text{ in at most } t(|w|) \text{ time} \,\},$$

and the *t-space-bounded Kolmogorov complexity* of $w$ is

$$\mathrm{KS}^t(w) = \min_{\pi \in \{0,1\}^*} \{\, |\pi| \mid U(\pi) = w \text{ in at most } t(|w|) \text{ space} \,\}.$$

**Fact 2.1.** *For all $w \in \{0,1\}^*$ and $t : \mathbb{N} \to \mathbb{N}$, $\mathrm{K}(w) \leq \mathrm{KS}^t(w) \leq \mathrm{K}^t(w)$.*

For all $q \in \mathbb{Q}$, let $\mathrm{K}(q) = \mathrm{K}(b_q)$, $\mathrm{K}^t(q) = \mathrm{K}^t(b_q)$ and $\mathrm{KS}^t(q) = \mathrm{KS}^t(b_q)$, where $b_q \in \{0,1\}^*$ is some standard binary representation of the rational $q$ with a numerator, denominator, and sign bit.

Define the *self-delimiting encoding function* $\mathrm{enc} : \{0,1\}^* \to \{0,1\}^*$ for all $w \in \{0,1\}^*$ by

$$\mathrm{enc}(w) = 0^{s_{|w|}} 1 s_{|w|} w.$$

For all $n \in \mathbb{N}$, let $\mathrm{enc}(n) = \mathrm{enc}(s_n)$. Strings encoded by enc and valid programs for $U$ are *self-delimiting.* They can be prepended to arbitrary strings and uniquely decoded.

**Observation 2.2.** *For all $w \in \{0,1\}^*$, $|\mathrm{enc}(w)| \leq |w| + 2\log|w| + 3$, and for all $n \in \mathbb{N}$ such that $n \geq 2$, $\mathrm{enc}(n) \leq \log n + 2\log\log n + 3$.*

Our results, being asymptotic in nature, do not depend crucially on using the self-delimiting Kolmogorov complexity K; it is simply more convenient for encoding purposes. All results in this paper hold if we use the plain Kolmogorov complexity $\mathrm{C} : \{0,1\}^* \to \mathbb{N}$ (see [30]) instead. Whenever we would need to add a program to a string and retain the ability to uniquely decode it, we could simply encode the program using the function enc.

## 2.3 Space/Time Bounds

The following explanation of growth rates and function classes is taken nearly verbatim from [31].

For each $i \in \mathbb{N}$, define a class $G_i$ of *growth rates* between linear and exponential as follows.

$$\begin{aligned} G_0 &= \{\ t : \mathbb{N} \to \mathbb{N} \mid (\exists k \in \mathbb{N})(\forall^\infty n \in \mathbb{N})\ t(n) \le kn\ \} \\ G_{i+1} &= 2^{G_i(\log n)} = \left\{\ t : \mathbb{N} \to \mathbb{N} \ \middle|\ (\exists g \in G_i)(\forall^\infty n \in \mathbb{N})\ t(n) \le 2^{g(\log n)}\ \right\} \end{aligned}$$

Unless stated otherwise, in this paper, for each $i \in \mathbb{N}$, $\Delta$ represents any of the following classes of functions

$$\begin{aligned} \mathrm{comp} &= \{\ f : \{0,1\}^* \to \{0,1\}^* \mid f \text{ is computable }\}, \\ \mathrm{p}_i &= \{\ f : \{0,1\}^* \to \{0,1\}^* \mid f \text{ is computable in } G_i \text{ time }\}, \\ \mathrm{p}_i\mathrm{space} &= \{\ f : \{0,1\}^* \to \{0,1\}^* \mid f \text{ is computable in } G_i \text{ space }\}. \end{aligned}$$

For example, $\mathrm{p}_0\mathrm{space}$ is the set of all functions computable in linear space, $\mathrm{p}_1\mathrm{space}$, abbreviated pspace, is the set of all functions computable in polynomial space, and $\mathrm{p}_2\mathrm{space}$ is the set of all functions computable in space $n^{(\log n)^k}$ for some $k \in \mathbb{N}$. Given a class of functions $\Delta$, let $\mathrm{bound}(\Delta) \subseteq \{t : \mathbb{N} \to \mathbb{N}\}$ denote the class of time or space bounds defining $\Delta$: bound(comp) is the set of all computable functions $t : \mathbb{N} \to \mathbb{N}$, and, for all $i \in \mathbb{N}$, $\mathrm{bound}(\mathrm{p}_i) = \mathrm{bound}(\mathrm{p}_i\mathrm{space}) = G_i$.

If $D$ is a discrete domain, we say a function $f : D \to \mathbb{R}$ is $\Delta$-*computable* if there is a function $\widehat{f} : D \times \mathbb{N} \to \mathbb{Q}$ such that $|\widehat{f}(x,r) - f(x)| \le 2^{-r}$ for all $r \in \mathbb{N}$ and $x \in D$, and $\widehat{f} \in \Delta$ (with $r$ coded in unary and $x$ and the output coded in binary). We say that $f$ is *exactly* $\Delta$-*computable* if $f : D \to \mathbb{Q}$ and $f \in \Delta$, and we say that $f$ is *dyadically* $\Delta$-*computable* if $f : D \to \mathbb{Q}_2$ and $f \in \Delta$.

## 2.4 Reductions

Let $M$ be a Turing machine and $S \in \mathbf{C}$. We say $M$ *computes* $S$ if, on input $n \in \mathbb{N}$, $M$ outputs the string $S \upharpoonright n$. We define an *oracle Turing machine* (*OTM*) to be a Turing machine $M$ that can make constant-time queries to an oracle sequence, and we let OTM denote the set of all oracle Turing machines. For $R \in \mathbf{C}$, we say $M$ operates *with oracle* $R$ if, whenever $M$ makes a query to index $n \in \mathbb{N}$, the bit $R[n]$ is returned. We write $M^R$ to denote the OTM $M$ with oracle $R$.

Let $S, R \in \mathbf{C}$ and $M \in$ OTM. We say *$S$ is Turing reducible to $R$ via $M$*, and we write $S \leq_{\mathrm{T}} R$ *via* $M$, if $M^R$ computes $S$.[1] In this case, define $M(R) = S$. We say *$S$ is Turing reducible to* $R$, and we write $S \leq_{\mathrm{T}} R$, if there exists $M \in$ OTM such that $S \leq_{\mathrm{T}} R$ via $M$. We say *$S$ is $\Delta$-Turing reducible to $R$ via $M$*, and we write $S \leq_{\mathrm{T}}^{\Delta} R$ *via* $M$, if $M^R$ computes $S$, and there is a function $q \in \mathrm{bound}(\Delta)$ such that, for all $n \in \mathbb{N}$, $M$ outputs $S \upharpoonright n$ using at most $q(n)$ time or space (depending on the resource defining $\Delta$). We say *$S$ is $\Delta$-Turing reducible to* $R$, and we write $S \leq_{\mathrm{T}}^{\Delta} R$, if there exists $M \in$ OTM such that $S \leq_{\mathrm{T}}^{\Delta} R$ via $M$. We say $S$ is *Turing equivalent* to $R$, and we write $S \equiv_{\mathrm{T}} R$, if $S \leq_{\mathrm{T}} R$ and $R \leq_{\mathrm{T}} S$, and we say $S$ is $\Delta$-*Turing equivalent* to $R$, and we write $S \equiv_{\mathrm{T}}^{\Delta} R$, if $S \leq_{\mathrm{T}}^{\Delta} R$ and $R \leq_{\mathrm{T}}^{\Delta} S$.

If $\Delta = \mathrm{comp}$, then a $\Delta$-Turing reduction is nothing more than a *truth-table reduction* (see [48]).[2] We write $S \leq_{\mathrm{tt}} R$ to denote that $S$ is truth-table reducible to $R$ (i.e., that $S \leq_{\mathrm{T}}^{\mathrm{comp}} R$). If $\Delta = \mathrm{p}_i\mathrm{space}$ or $\mathrm{p}_i$, and we identify a sequence $S \in \mathbf{C}$ with the language $L \subseteq \{0,1\}^*$ for which $S = \chi_L$, then a $\Delta$-Turing reduction is an $\mathsf{E}_i\mathsf{SPACE}$ or $\mathsf{E}_i$ (see [31]) Turing reduction, respectively. Since this paper deals exclusively with sequences, we will use the convention of calling such a reduction a $\mathrm{p}_i$space- or $\mathrm{p}_i$-Turing reduction, indicating that the polynomial bound is in terms of the length of the prefix of the characteristic sequence (the output) and not in terms of the length of the strings in the language (the input).

Let $S, P, R \in \mathbf{C}$ and $M_S, M_P \in$ OTM such that $S \leq_{\mathrm{T}} P$ via $M_S$ and $P \leq_{\mathrm{T}} R$ via $M_P$. Define the *composition of $M_S$ with $M_P$*, denoted $M_S \circ M_P$, to be the OTM that works as follows. On input $n \in \mathbb{N}$ and with oracle $R$, $(M_S \circ M_P)^R$ simulates $M_S^P$ to compute $S \upharpoonright n$. Whenever a bit of $P$ is queried by $M_S$, $(M_S \circ M_P)^R$ simulates $M_P^R$ for the minimum number of steps needed to compute that bit of $P$.

**Observation 2.3.** *$\leq_{\mathrm{T}}$ is transitive: if $S \leq_{\mathrm{T}} P$ via $M_S$ and $P \leq_{\mathrm{T}} R$ via $M_P$, then $S \leq_{\mathrm{T}} R$ via $M_S \circ M_P$.*

# 3 Decompression and Dimension

## 3.1 Decompression via Reductions

Let $S, R \in \mathbf{C}$ and $M \in$ OTM such that $S \leq_{\mathrm{T}} R$ via $M$. Define $\#(S \upharpoonright n, M^R)$ to be the *query usage of $M^R$ on $S \upharpoonright n$*, the number of bits of $R$ queried by $M$ when computing the string $S \upharpoonright n$. (If we instead define $\#(S \upharpoonright n, M^R)$ to be the index of the rightmost bit of $R$

[1] This differs from the more standard definition of a Turing reduction in that $S \upharpoonright n$, instead of $S[n]$, is computed on input $n$. The definitions are equivalent if at least polynomial time is allowed.

[2] A truth-table reduction is typically defined to be a Turing reduction that halts on all oracles; it is easy to to see that this occurs if and only if the time (equivalently, space) used by the reduction on input $n$ is bounded by a computable function of $n$.

queried by $M$ when computing $S \upharpoonright n$, all results of the present paper still hold.) Define

$$\rho_M^-(S,R) = \liminf_{n\to\infty} \frac{\#(S \upharpoonright n, M^R)}{n},$$
$$\rho_M^+(S,R) = \limsup_{n\to\infty} \frac{\#(S \upharpoonright n, M^R)}{n}.$$

Viewing $R$ as a compressed version of $S$, $\rho_M^-(S,R)$ and $\rho_M^+(S,R)$ are respectively the best- and worst-case compression ratios as $M$ decompresses $R$ into $S$. Note that $0 \le \rho_M^-(S,R) \le \rho_M^+(S,R) \le \infty$. For $S \in \mathbf{C}$, the *lower and upper Turing decompression ratios of $S$* are respectively defined

$$\rho^-(S) = \min_{\substack{R\in\mathbf{C}\\ M\in\mathrm{OTM}}} \left\{ \rho_M^-(S,R) \;\middle|\; S \le_{\mathrm{T}} R \text{ via } M \right\},$$
$$\rho^+(S) = \min_{\substack{R\in\mathbf{C}\\ M\in\mathrm{OTM}}} \left\{ \rho_M^+(S,R) \;\middle|\; S \le_{\mathrm{T}} R \text{ via } M \right\}.$$

Note that $0 \le \rho^-(S) \le \rho^+(S) \le 1$. As we will see, by Lemma 5.2 and Theorem 5.5, the two minima above exist. In fact, there is a single OTM $M$ that achieves the minimum decompression ratio in each case.

The *lower and upper $\Delta$-Turing decompression ratios of $S$* are respectively defined

$$\rho_\Delta^-(S) = \inf_{\substack{R\in\mathbf{C}\\ M\in\mathrm{OTM}}} \left\{ \rho_M^-(S,R) \;\middle|\; S \le_{\mathrm{T}}^{\Delta} R \text{ via } M \right\},$$
$$\rho_\Delta^+(S) = \inf_{\substack{R\in\mathbf{C}\\ M\in\mathrm{OTM}}} \left\{ \rho_M^+(S,R) \;\middle|\; S \le_{\mathrm{T}}^{\Delta} R \text{ via } M \right\}.$$

Recall that a $\le_{\mathrm{T}}^{\mathrm{comp}}$-reduction is simply a truth-table reduction. Therefore, for all $S \in \mathbf{C}$, the *lower and upper truth-table decompression ratios* of $S$ are respectively defined

$$\rho_{\mathrm{tt}}^-(S) = \rho_{\mathrm{comp}}^-(S) = \inf_{\substack{R\in\mathbf{C}\\ M\in\mathrm{OTM}}} \left\{ \rho_M^-(S,R) \;\middle|\; S \le_{\mathrm{tt}} R \text{ via } M \right\},$$
$$\rho_{\mathrm{tt}}^+(S) = \rho_{\mathrm{comp}}^+(S) = \inf_{\substack{R\in\mathbf{C}\\ M\in\mathrm{OTM}}} \left\{ \rho_M^+(S,R) \;\middle|\; S \le_{\mathrm{tt}} R \text{ via } M \right\}.$$

## 3.2 Effective Dimension

See [33, 34] for an introduction to the theory of effective dimension.

Effective dimension was first defined in [33]. It is based on martingales, which are strategies for betting on bits of an infinite sequence.

1. An *$s$-gale* is a function $d : \{0,1\}^* \to [0,\infty)$ such that, for all $w \in \{0,1\}^*$,

$$d(w) = 2^{-s}[d(w0) + d(w1)].$$

2. A *martingale* is a 1-gale.

Intuitively, a martingale is a strategy for gambling in the following game. The gambler starts with some initial amount of *capital* (money) $d(\lambda)$, and it reads an infinite sequence $S$ of bits. $d(w)$ represents the capital the gambler has after reading the prefix $w \sqsubset S$. Based on $w$, the gambler bets some fraction of its capital that the next bit will be 0 and the remainder of its capital that the next bit will be 1. The capital bet on the bit that appears next is doubled, and the remaining capital is lost. The condition $d(w) = \frac{d(w0)+d(w1)}{2}$ ensures *fairness*: the martingale's expected capital after seeing the next bit, given that it has already seen the string $w$, is equal to its current capital. The fairness condition and an easy induction lead to the following observation.

**Observation 3.1.** *Let $k \in \mathbb{N}$ and let $d : \{0,1\}^* \to [0,\infty)$ be a martingale. Then*

$$\sum_{u \in \{0,1\}^k} d(u) = 2^k d(\lambda).$$

An $s$-gale is a martingale in which the capital bet on the bit that occurred is multiplied by $2^s$, as opposed to simply 2, after each bit. The parameter $s$ may be regarded as the *unfairness of the betting environment*; the lower the value of $s$, the faster money is taken away from the gambler. Let $d : \{0,1\}^* \to [0,\infty)$ be a martingale and let $s \in [0,\infty)$. Define the *$s$-gale induced by $d$*, denoted $d^{(s)}$, for all $w \in \{0,1\}^*$ by

$$d^{(s)}(w) = 2^{(s-1)|w|} d(w).$$

If a gambler's martingale is given by $d$, then, for all $s \in [0,\infty)$, its $s$-gale is $d^{(s)}$.

The following theorem, due to Lutz, establishes an upper bound on the number of strings on which an $s$-gale can perform well.

**Theorem 3.2** ( [33]). *Let $d$ be an $s$-gale. Then for all $w \in \{0,1\}^*$, $k \in \mathbb{N}$, and $\alpha \in \mathbb{R}^+$, there are fewer than $\frac{2^k}{\alpha}$ strings $u \in \{0,1\}^k$ for which*

$$\max_{v \sqsubseteq u} \left\{ 2^{(1-s)|v|} d(wv) \right\} \geq \alpha d(w).$$

**Corollary 3.3.** *Let $d$ be a martingale. Then for all $l \in \mathbb{R}$, $w \in \{0,1\}^*$, $k \in \mathbb{N}$, and $\alpha \in \mathbb{R}^+$, there are fewer than $\frac{2^l}{\alpha}$ strings $u \in \{0,1\}^k$ for which*

$$d(wu) \geq \alpha 2^{k-l} d(w).$$

Let $S \in \mathbf{C}$, $s \in [0,\infty)$, and let $d : \{0,1\}^* \to [0,\infty)$ be an $s$-gale. $d$ *succeeds* on $S$, and we write $S \in \mathrm{S}^\infty[d]$, if

$$\limsup_{n\to\infty} d(S \upharpoonright n) = \infty.$$

$d$ *strongly succeeds* on $S$, and we write $S \in \mathrm{S}^\infty_{\mathrm{str}}[d]$, if

$$\liminf_{n\to\infty} d(S \upharpoonright n) = \infty.$$

The next lemma follows easily from the proof of the Exact Computation Lemma of [33].

**Lemma 3.4** ( [33]). *If $d$ is a $\Delta$-computable $s$-gale and $2^s$ is a dyadic rational, then there is a dyadically $\Delta$-computable $s$-gale $\widetilde{d}$ such that $S^\infty[d] \subseteq S^\infty[\widetilde{d}]$ and $S^\infty_{\mathrm{str}}[d] \subseteq S^\infty_{\mathrm{str}}[\widetilde{d}]$.*

Let $d : \{0,1\}^* \to [0,\infty)$ be an $s$-gale. We say that $d$ is *constructive (a.k.a. lower semicomputable, subcomputable)* if there is a computable function $\widehat{d} : \{0,1\}^* \times \mathbb{N} \to \mathbb{Q}$ such that, for all $w \in \{0,1\}^*$ and $t \in \mathbb{N}$,

1. $\widehat{d}(w,t) \leq \widehat{d}(w,t+1) < d(w)$, and
2. $\lim\limits_{t\to\infty} \widehat{d}(w,t) = d(w)$.

Let $R \in \mathbf{C}$. We say that $R$ is *Martin-Löf random*, and we write $R \in \mathsf{RAND}$, if there is no constructive martingale $d$ such that $R \in S^\infty[d]$. This characterization of Martin-Löf randomness, due to Schnorr [44], is equivalent to Martin-Löf's traditional definition (see [30, 35]).

If there is a martingale $d$ that succeeds on a sequence $S \in \mathbf{C}$, then $d$ makes arbitrarily high capital on $S$. Using a standard technique (cf. [37]), one may construct from $d$ a martingale $d'$ that *strongly* succeeds on $S$. This is done by maintaining a "side account" of capital that is not used to bet: i.e., the capital in that account is always allocated equally between 0 and 1 when betting. Whenever $d$ makes strictly more than \$1, \$1 is moved into the side account. Since $d$ succeeds on $S$, it will eventually make more than \$1 in the main account again, and so infinitely often, the side account will grow by \$1, whence $d'$ strongly succeeds on $S$. It is clear that if $d$ is $\Delta$-computable, then $d'$ is also $\Delta$-computable.

**Observation 3.5.** *Let $S \in \mathbf{C}$ such that there is a $\Delta$-computable martingale that succeeds on $S$. Then there is a $\Delta$-computable martingale that strongly succeeds on $S$.*

This technique works for constructive dimension as well. Furthermore, Schnorr [44] showed that there is a universal constructive martingale that succeeds on *every* sequence not in $\mathsf{RAND}$. This leads to the following observation.

**Observation 3.6.** *There is a constructive martingale $\mathbf{d}$ such that $S^\infty_{\mathrm{str}}[\mathbf{d}] = \mathsf{RAND}^{\mathrm{c}}$.*

Let $\widehat{\mathbf{d}} : \{0,1\}^* \times \mathbb{N} \to \mathbb{Q}$ be the computable function testifying that $\mathbf{d}$ is constructive.

The following theorem, due independently to Hitchcock and Fenner, states that $\mathbf{d}^{(s)}$ is "optimal" for the class of constructive $t$-gales whenever $s > t$.

**Theorem 3.7** ( [16, 24]). *Let $s > t \in [0,\infty)$, and let $d$ be a constructive $t$-gale. Then*

$$S^\infty[d] \subseteq S^\infty[\mathbf{d}^{(s)}] \text{ and } S^\infty_{\mathrm{str}}[d] \subseteq S^\infty_{\mathrm{str}}[\mathbf{d}^{(s)}].$$

By Theorem 3.7, the following definition of constructive dimension is equivalent to the definitions given in [2, 34]. Let $S \in \mathbf{C}$. The *(constructive) dimension* and the *(constructive) strong dimension* of $S$ are respectively defined

$$\begin{aligned}\dim(S) &= \inf\{s \in [0,\infty) \mid S \in S^\infty[\mathbf{d}^{(s)}]\},\\ \mathrm{Dim}(S) &= \inf\{s \in [0,\infty) \mid S \in S^\infty_{\mathrm{str}}[\mathbf{d}^{(s)}]\}.\end{aligned}$$

Intuitively, the constructive (strong) dimension of $S$ is the *most unfair betting environment* $s$ in which the optimal constructive gambler $\mathbf{d}$ (strongly) succeeds on $S$.

**Observation 3.8.** *If* $S \in \mathsf{RAND}$*, then* $\dim(S) = \mathrm{Dim}(S) = 1$.

The following theorem, due to Mayordomo [36], gives a useful characterization of the constructive dimension of a sequence in terms of Kolmogorov complexity, and it justifies the intuition that constructive dimension measures the *density of algorithmic information* in a sequence.

**Theorem 3.9** ( [36])**.** *For all* $S \in \mathbf{C}$*,*

$$\dim(S) = \liminf_{n\to\infty} \frac{\mathrm{K}(S \upharpoonright n)}{n},$$

*and*

$$\mathrm{Dim}(S) = \limsup_{n\to\infty} \frac{\mathrm{K}(S \upharpoonright n)}{n}.$$

Let $G_\Delta^{(s)}$ denote the set of all $\Delta$-computable $s$-gales. For all $S \in \mathbf{C}$, the $\Delta$-*dimension* and the $\Delta$-*strong dimension* of $S$ are respectively defined

$$\dim_\Delta(S) = \inf\left\{ s \in [0,\infty) \;\middle|\; \left(\exists d \in G_\Delta^{(s)}\right)\ S \in \mathrm{S}^\infty[d] \right\},$$

and

$$\mathrm{Dim}_\Delta(S) = \inf\left\{ s \in [0,\infty) \;\middle|\; \left(\exists d \in G_\Delta^{(s)}\right)\ S \in \mathrm{S}^\infty_{\mathrm{str}}[d] \right\}.$$

We say $S \in \mathbf{C}$ is $\Delta$-*regular* if $\dim_\Delta(S) = \mathrm{Dim}_\Delta(S)$.

The following alternate characterization of the space-bounded $\Delta$-dimensions, resembling Theorem 3.9, is due to Hitchcock [22].

**Theorem 3.10** ( [22])**.** *Let* $i \in \mathbb{N}$ *and* $\Delta \in \{\mathrm{comp}, \mathrm{p}_i\mathrm{space}\}$*. For all* $S \in \mathbf{C}$*,*

$$\dim_\Delta(S) = \inf_{p\in\mathrm{bound}(\Delta)} \liminf_{n\to\infty} \frac{\mathrm{KS}^p(S \upharpoonright n)}{n},$$

*and*

$$\mathrm{Dim}_\Delta(S) = \inf_{p\in\mathrm{bound}(\Delta)} \limsup_{n\to\infty} \frac{\mathrm{KS}^p(S \upharpoonright n)}{n}.$$

Let $r, t \in [0,\infty)$ and $S \in \mathbf{C}$. Note that if $d_1$ and $d_2$ are martingales such that $d_1^{(t)}$ succeeds on $S$ and $d_2^{(r)}$ strongly succeeds on $S$, then $d_1$ and $d_2$ can be combined into a single martingale $d$ that simulates $d_1$ and $d_2$ in separate "accounts". Furthermore, if $d_1$ and $d_2$ are both $\Delta$-computable, then $d$ is also $\Delta$-computable. This leads to the following observation.

**Observation 3.11.** *Let* $S \in \mathbf{C}$*,* $t > \dim_\Delta(S)$*, and* $r > \mathrm{Dim}_\Delta(S)$*. Then there is a* $\Delta$*-computable martingale* $d$ *such that* $d^{(t)}$ *succeeds on* $S$ *and* $d^{(r)}$ *strongly succeeds on* $S$*.*

If $d$ is a $\Delta$-computable $s$-gale, then $d' : \{0,1\}^* \to [0,\infty)$ defined for all $w \in \{0,1\}^*$ by $d'(w) = d(w)/d(\lambda)$ is also an $s$-gale. It is clear that $d'$ is $\Delta$-computable, and that $d'$ (strongly) succeeds on the same sequences on which $d$ (strongly) succeeds, whence the following holds.

**Observation 3.12.** *Let $s > 0$. For every $\Delta$-computable $s$-gale $d$, there exists a $\Delta$-computable $s$-gale $d'$ such that $d'(\lambda) = 1$, $\mathrm{S}^\infty[d'] = \mathrm{S}^\infty[d]$, and $\mathrm{S}^\infty_{\mathrm{str}}[d'] = \mathrm{S}^\infty_{\mathrm{str}}[d]$.*

The same technique cannot be used in general for a constructive $s$-gale $d$, since the initial capital $d(\lambda)$ may be uncomputable. However, by dividing each value $d(w)$ by a rational approximation to $d(\lambda)$, we may assume that all constructive $s$-gales have initial capital arbitrarily close to 1.

**Observation 3.13.** *Let $s > 0$ and $\epsilon > 0$. For every constructive $s$-gale $d$, there exists a constructive $s$-gale $d'$ such that $1 \leq d'(\lambda) < 1 + \epsilon$, $\mathrm{S}^\infty[d'] = \mathrm{S}^\infty[d]$, and $\mathrm{S}^\infty_{\mathrm{str}}[d'] = \mathrm{S}^\infty_{\mathrm{str}}[d]$.*

Though an analogue of Theorem 3.10 for $\mathrm{p}_i$-bounded Kolmogorov complexity and the $\mathrm{p}_i$-dimensions is currently unknown, López-Valdés and Mayordomo [51] have proved a characterization of polynomial-time dimension and strong dimension in terms of *reversible* polynomial-time compressors and decompressors. This characterization will be useful in constructing time-bounded dimension extractors. It is easily verified that their characterization extends to the $\mathrm{p}_i$-dimensions for $i \geq 1$.

López-Valdés and Mayordomo characterize the $\mathrm{p}_i$-dimensions as follows. Let $i \in \mathbb{Z}^+$. A $\mathrm{p}_i$-*compressor* is a pair $(C, D)$ of functions $C : \{0,1\}^* \to \{0,1\}^*$ and $D : \{0,1\}^* \times \mathbb{N} \to \{0,1\}^*$, each computable in $\mathrm{p}_i$ time, such that, for all $w \in \{0,1\}^*$, $D(C(w), |w|) = w$. A $\mathrm{p}_i$-compressor *does not start from scratch* if, for all but finitely many $k \in \mathbb{N}$ and $w \in \{0,1\}^*$,

$$\sum_{u \in \{0,1\}^k} 2^{-|C(wu)|} \leq 2^{k/\log k} 2^{-|C(w)|}.$$

In particular, López-Valdés and Mayordomo note that a $\mathrm{p}_i$-compressor $(C, D)$ does not start from scratch if, for all $w, u \in \{0,1\}^*$, $C(w)$ and $C(wu)$ have a common prefix of length at least

$$|C(w)| - \frac{|u|}{\log |u|} + \log |u|. \tag{3.1}$$

Let $\mathcal{PC}_i$ denote the set of all $\mathrm{p}_i$-compressors that do not start from scratch.

**Theorem 3.14** ( [51])**.** *For all $i \in \mathbb{Z}^+$ and $S \in \mathbf{C}$,*

$$\dim_{\mathrm{p}_i}(S) = \inf \left\{ (C, D) \in \mathcal{PC}_i \;\middle|\; \liminf_{n\to\infty} \frac{|C(S \upharpoonright n)|}{n} \right\},$$

*and*

$$\mathrm{Dim}_{\mathrm{p}_i}(S) = \inf \left\{ (C, D) \in \mathcal{PC}_i \;\middle|\; \limsup_{n\to\infty} \frac{|C(S \upharpoonright n)|}{n} \right\}.$$

# 4 Dimension Extractors

This section investigates various effective dimensions in which effective reducibilities may be used as extractors: constructive dimension, extracted with Turing reductions, computable dimension, extracted with truth-table reductions, $\mathrm{p}_i$space-dimension, extracted with $\mathrm{p}_i$space-bounded Turing reductions, $\mathrm{p}_i$-dimension, extracted with $\mathrm{p}_i$-bounded Turing reductions, and finite-state dimension, extracted with information lossless finite-state transducers.

To understand the motivation for studying dimension extractors, it is helpful to understand classical extractors. In general, an extractor is an algorithm used to transform a source of weak randomness into a source of stronger randomness. Extractors are motivated in part by the abundance of weak random sources in nature – for instance, electrical noise from Zener diodes – and the need for uniform (i.e., strong) random sources in probabilistic algorithms. Von Neumann's [53] coin flip technique is the simplest and most famous extractor: a biased coin may be used to simulate an unbiased coin by always flipping the coin twice, ignoring the combinations HH and TT, and interpreting HT to mean H and TH to mean T.

More formally, in computational complexity theory, a extractor is a function, generally computable in polynomial time, taking as input a string drawn from a probability distribution $X$ on $\{0,1\}^n$ with *min-entropy* at least $k$, and a much smaller string of length $d$, called the *seed*, drawn from the uniform distribution on $\{0,1\}^d$. The extractor's output is "close" to uniformly distributed, but much longer than the seed. The min-entropy of $X$ is defined $\min_{x\in\{0,1\}^n} \log \Pr[x = X]^{-1}$; it is the Shannon self-information [47] of the string with the highest probability in $\{0,1\}^n$. If $k$ is strictly between 0 and $n$, $X$ may be thought of as "partially random"; $n$ bits drawn from $X$ have at least $k$ bits of randomness. The goal of an extractor is to transform $X$ into a distribution that is closer to "fully random", i.e., to output $m$ bits that have close to $m$ bits of randomness. See [46] for a comprehensive survey of extractors in computational complexity.

For algorithmic purposes, a deterministic infinite sequence that *appears* random to any algorithm often works just as well as a truly probabilistic source. The complexity class $\mathsf{BPP}$, defined by Gill [19] to be those languages decidable by a randomized polynomial time algorithm with probability of correctness at least 2/3, is generally regarded as the set of decision problems feasibly decidable by a randomized algorithm. Bennett [4] (refining measure-theoretic arguments of [5] and [1]) has demonstrated that, given access to *any* oracle sequence that is algorithmically random in the sense of Martin-Löf, every language in $\mathsf{BPP}$ can be decided *deterministically* in polynomial time. Book, Lutz, and Wagner [8] have shown a wealth of similar characterizations of $\mathsf{BPP}$ and other randomized complexity classes in terms of oracle access to Martin-Löf random sequences. Lutz [32], using the techniques of resource-bounded measure [31], improved the result of Bennett by showing that all of $\mathsf{BPP}$ can be decided in polynomial time relative to any *pspace-random* oracle, which is a sequence that appears random to any polynomial-space-bounded algorithm.

Recall that, given a resource bound $\Delta$, $\Delta$-dimension quantifies how close a sequence is to being $\Delta$-random. Gu and Lutz [20] improved Lutz's above-mentioned pspace-random oracle result by showing that all of BPP is polynomial time decidable relative to any oracle sequence with positive pspace-dimension (in fact, the oracle needs only have positive $\Delta_3^{\mathrm{p}}$-dimension). Therefore, for certain applications, if a sequence has positive effective dimension, it contains sufficient *algorithmic* randomness to act as a substitute for a truly probabilistic source. This highlights the parallels between the theory of effective dimension and randomness extractors. A non-random sequence with positive dimension may be considered "weakly random": the first $n$ bits of a sequence with Hausdorff and packing $\Delta$-dimension equal to $\alpha$ contain about $\alpha n$ bits of $\Delta$-randomness.

Reimann [41] and Terwijn asked the question, given any sequence $S$ such that $\dim(S) > 0$, does oracle access to $S$ allow us to compute a Martin-Löf random sequence? Miller and Nies [38] posed the related questions, does oracle access to $S$ allow us to compute a sequence of constructive dimension 1, or arbitrarily close to 1, or strictly greater than $\dim(S)$? Viewing constructive dimension as a quantification of the amount of randomness contained in a sequence, a computation increasing the constructive dimension of a sequence performs the same function as the extractors mentioned earlier: the computation transforms a partially random source into a more random source. Therefore Reimann, Terwijn, Miller and Nies are essentially asking whether Turing reductions can extract constructive dimension.

## 4.1 Constructive Dimension Extractors

We show that constructive dimension can be extracted in a weaker sense. Using Ryabko's result [42, 43] on optimal reversible compression of sequences, we demonstrate that, for every $\epsilon > 0$ and every sequence $S$ such that $\dim(S) > 0$, there is a sequence $P$, Turing equivalent to $S$, such that $\mathrm{Dim}(P) \geq 1 - \epsilon$. In fact, there is a single oracle Turing machine that accomplishes this extraction, taking a rational $\beta > \dim(S)$ as an input parameter used to control the size of $\epsilon$. Moreover, the extractor uses close to an optimal number of bits of the input sequence to compute the output sequence, in the sense that for infinitely many $n$, about $n$ bits are required from $S$ to compute $\dim(S) \cdot n$ bits of $P$.

The next theorem is due to Ryabko [42, 43].

**Theorem 4.1** ( [42, 43]). *There exist OTMs $M_e$ and $M_d$, with $M_e$ taking a single input $\beta \in \mathbb{Q}$, with the property that, for every $S \in \mathbf{C}$ and every rational $\beta > \dim(S)$, there exists $P \in \mathbf{C}$ such that $P \leq_{\mathrm{T}} S$ via $M_e(\beta)$, $S \leq_{\mathrm{T}} P$ via $M_d$, and $\rho^-_{M_d}(S, P) < \beta$.*

The next theorem states that any sequence in which almost every prefix has Kolmogorov rate bounded away from zero can be used to compute a sequence with infinitely many prefixes of nearly maximal Kolmogorov rate. Furthermore, this can be done with a single OTM taking a rational upper bound on the constructive dimension of the input sequence.

**Theorem 4.2.** *There exists an OTM $M_e$, with $M_e$ taking $\beta \in \mathbb{Q}$ as input, such that, for all $S \in \mathbf{C}$ such that $\dim(S) > 0$, and all $\epsilon > 0$ such that $0 < \beta - \dim(S) \leq \epsilon \cdot \dim(S)/3$, there exists $P \in \mathbf{C}$ such that $P \leq_{\mathrm{T}} S$ via $M_e(\beta)$, $S \leq_{\mathrm{T}} P$, and $\mathrm{Dim}(P) \geq 1 - \epsilon$.*

Laurent Bienvenu [6] has shown that the bound of $\beta - \dim(S) \leq \epsilon \cdot \dim(S)/3$ required in the hypothesis of Theorem 4.2 can be improved to $\beta - \dim(S) \leq \epsilon \cdot \dim(S)$.

The statement of Theorem 4.2 is complicated by the rational input $\beta$ required to make the OTM $M_e$ uniform over all sequences. The following corollary states simply that Turing reductions can extract constructive strong dimension from positive constructive dimension.

**Corollary 4.3.** *For each* $\epsilon > 0$ *and each* $S \in \mathbf{C}$ *such that* $\dim(S) > 0$*, there exists* $P \in \mathbf{C}$ *such that* $P \equiv_{\mathrm{T}} S$ *and* $\mathrm{Dim}(P) \geq 1 - \epsilon$*.*

*Proof of Theorem 4.2.* Let $S$, $\beta$, and $\epsilon$ be as in the statement of the theorem, and define $\delta = \beta - \dim(S) > 0$. Let $M_e, M_d \in \mathrm{OTM}$ be as in Theorem 4.1, so that $M_e^S(\beta)$ computes $P$, $M_d^P$ computes $S$, and, letting $n_p = \#(S \restriction n, M_d^P)$ for all $i \in \mathbb{N}$, since $\rho^-_{M_d}(S, P) < \beta$,

$$
\begin{aligned}
& \liminf_{n\to\infty} \frac{n_p}{n} < \beta \\
\Longrightarrow \quad & (\exists^\infty n \in \mathbb{N}) \ \liminf_{m\to\infty} \frac{\mathrm{K}(S \restriction m)}{m} > \frac{n_p}{n} - \delta \quad \text{by Theorem 3.9} \\
\Longrightarrow \quad & (\exists^\infty n \in \mathbb{N}) \ \mathrm{K}(S \restriction n) > n_p - \delta n. \qquad (4.1)
\end{aligned}
$$

Since $\delta \leq \epsilon \cdot \dim(S)/3 < \epsilon \cdot \dim(S)/2$,

$$(\forall^\infty n \in \mathbb{N}) \ \mathrm{K}(S \restriction n) > \frac{2}{\epsilon}\delta n. \qquad (4.2)$$

Ryabko's construction of $M_d$ is such that entire prefixes of the oracle sequence are queried at once: whenever the bit at index $i \in \mathbb{N}$ is queried, all bits $j < i$ are also queried. Thus, a program $M$ simulating $M_d$ with the first $n_p$ bits of $P$ can calculate $S \restriction n$, and $M$ can be encoded in $|\mathrm{enc}(P \restriction n_p)| + O(1)$ bits. Thus there is a constant $c$ such that $\mathrm{K}(S \restriction n) \leq n_p + 2\log n_p + c$, which together with (4.2) implies that

$$(\forall^\infty n \in \mathbb{N}) \ \delta n < \frac{\epsilon}{2}(n_p + 2\log n_p + c). \qquad (4.3)$$

Combining (4.1) and (4.3),

$$(\exists^\infty n \in \mathbb{N}) \ \mathrm{K}(S \restriction n) > n_p - \frac{\epsilon}{2}\left(p_n + 2\log n_p + c\right). \qquad (4.4)$$

If the OTMs $M_e$ and $M_d$ are defined as in Theorem 4.1, then $M_e(\beta)$ and $M_d$ testify that $P \equiv_{\mathrm{T}} S$. It remains to show that $\mathrm{Dim}(P) \geq 1 - \epsilon$. Suppose for the sake of contradiction that $\mathrm{Dim}(P) < 1 - \epsilon$. Then it would be the case that

$$(\forall^\infty m \in \mathbb{N}) \ \mathrm{K}(P \restriction m) < m - \epsilon m. \qquad (4.5)$$

Since $\dim(S) > 0$, $S$ is uncomputable, and therefore $n_p$ grows unboundedly with $n$. A program that produces $P \restriction n_p$ can be used in conjunction with $M_d$ to produce $S \restriction n$. Therefore, for a suitable constant $c' \approx |M_d|$,

$$
\begin{aligned}
(\forall^\infty n \in \mathbb{N}) \ \mathrm{K}(S \restriction n) \ &\leq \ \mathrm{K}(P \restriction n_p) + c' \\
&< \ n_p - \epsilon n_p + c' \\
&< \ n_p - \frac{\epsilon}{2}\left(p_n + 2\log n_p + c\right). \qquad (4.6)
\end{aligned}
$$

But (4.6) contradicts (4.4). Hence, $\mathrm{Dim}(P) \geq 1 - \epsilon$. $\square$

Unfortunately, technique of the preceding proof does not show that Turing reductions are able to increase constructive dimension in addition to constructive strong dimension (i.e., that *almost every* prefix of the output sequence has high Kolmogorov complexity). Nies and Reimann [39] have shown that constructive dimension *cannot* be extracted with weak truth-table reductions (a Turing reduction in which the query usage on input $n$ is bounded by a computable function of $n$): for every rational $\alpha \in [0,1]$, there is a sequence $S$ such that $\dim(S) = \alpha$, and every sequence $P$ that is weak truth-table reducible to $S$ satisfies $\dim(P) \leq \alpha$. Since the Turing reduction in our proof is also a weak truth-table reduction, it cannot be the case that $\dim(P) > \dim(S)$.

Bienvenu, Doty, and Stephan [7] have shown that the stronger results of the next subsection hold for constructive dimension as well, using weak truth-table reductions, thereby improving the present paper's Corollary 4.3 to be as strong as Theorem 4.8 (and in fact stronger, as a *uniform* extractor is exhibited similar to that in Theorem 4.2). However, it is also shown in [7] that there is no uniform Turing reduction is capable of always increasing constructive dimension by a fixed amount: for every $\alpha, \beta$ with $0 < \alpha < \beta < 1$, and every Turing reduction $M$, there is a sequence $S$ such that $\dim(S) \geq \alpha$ and, if $M^S$ computes the sequence $R$, then $\dim(R) < \beta$. This implies that the reduction techniques of [7] and of the present paper, which use no property of the input sequence other than a simple bound on its dimension, cannot be used to prove the existence of constructive dimension extractors.

Buhrman, Fortnow, Newman, and Vereshchagin [10] and Fortnow, Hitchcock, Pavan, Vinodchandran, and Wang [17] have demonstrated related constructions for extracting Kolmogorov complexity from finite strings. Buhrman, Fortnow, Newman, and Vereshchagin show that there is an efficient algorithm, taking as input any non-random string, that outputs a small list of strings of the same length as the input string, where at least one output string is guaranteed to have higher Kolmogorov complexity than the input. Note that given a finite string $x$ and the value $\mathrm{K}(x)$, an algorithmically random string containing exactly the amount of algorithmic information in $x$ may be extracted from $x$: namely, a shortest program for $x$. The value $\mathrm{K}(x)$ – requiring at most $\log|x|$ bits to represent – may be considered "advice" bits that help the algorithm extract randomness from $x$. Fortnow, Hitchcock, Pavan, Vinodchandran, and Wang improve upon this observation by showing that there is an efficient algorithm such that, for any $\alpha, \beta$ such that $0 < \alpha < \beta < 1$, if the input string $x$ has Kolmogorov complexity at least $\alpha|x|$, then, given a *constant* (with respect to $\alpha$ and $\beta$) number of advice bits, the output string $y$ (with $|y| = \Omega(|x|)$) will have Kolmogorov complexity at least $\beta|y|$. This is shown to hold for space-bounded Kolmogorov complexity as well. The advice bits are necessary; Vereshchagin and Vyugin [52] have shown that no uniform algorithm is capable of extracting Kolmogorov complexity from finite strings.

## 4.2 Other Effective Dimension Extractors

As noted, Ryabko's result leads easily to an extractor for constructive dimension. In contrast, our main extractor result, Theorem 4.8, relies on new techniques to exhibit extractors for resource-bounded $\Delta$-dimension [33], where $\Delta$ represents the class comp of computable functions or, for each $i \in \mathbb{N}$, any of the classes $\mathrm{p}_i$space of $\mathrm{p}_i$space-computable functions or the classes $\mathrm{p}_i$ of $\mathrm{p}_i$-time computable functions. We show that for every $\epsilon > 0$ and every sequence $S$ such that $\dim_\Delta(S) > 0$, there is a sequence $P$, $\Delta$-Turing equivalent to $S$, such that $\mathrm{Dim}_\Delta(P) \geq 1 - \epsilon$. We show a similar result for finite-state dimension, with the extractor implemented by an information lossless finite-state transducer [26, 47]. In contrast to the constructive dimension extractor, a different reduction machine is required for each sequence $S$.

In addition to the near-optimal extraction of strong dimension, the extractors for computable, $\mathrm{p}_i$space, $\mathrm{p}_i$, and finite-state dimension are shown to partially extract dimension as well. More precisely, if the input sequence has dimension $d$ and strong dimension $D$, then the sequence output by the extractor has dimension at least $d/D - \epsilon$, where $\epsilon$ is an arbitrarily small positive constant. Therefore, for any *regular* sequence – a sequence whose dimension and strong dimension agree – dimension is nearly optimally extracted, in addition to strong dimension.

### 4.2.1 $\Delta$-Dimension Extractors

This subsection examines dimension extractors for computable dimension, space-bounded dimension, and time-bounded dimension.

An OTM that computes a sequence $S$, together with a finite prefix of the oracle that it queries, is a program to produce a prefix of $S$. Thus, the query usage of a space-bounded OTM on that prefix of $S$ cannot be far below the space-bounded Kolmogorov complexity of the prefix of $S$. This is formalized in the following lemma, which bounds the optimal compression ratio below by dimension.

**Lemma 4.4.** *Let $i \in \mathbb{N}$ and $\Delta \in \{\mathrm{comp}, \mathrm{p}_i\mathrm{space}\}$. For all $S \in \mathbf{C}$,*

$$\rho^-_\Delta(S) \geq \dim_\Delta(S), \text{ and } \rho^+_\Delta(S) \geq \mathrm{Dim}_\Delta(S).$$

*Proof.* Let $S, P \in \mathbf{C}$, and let $M \in \mathrm{OTM}$ such that $S \leq^\Delta_\mathrm{T} P$ via $M$. For $P = S$, $S \leq^\Delta_\mathrm{T} P$ via the trivial "bit-copier" OTM that always queries exactly $n$ bits of $P$ to compute $n$ bits of $S$, so we may assume that for all $n \in \mathbb{N}$, $\#(S \upharpoonright n, M^P) \leq n$. Thus, since $M$ has available at least a linear amount of space, we may assume that each bit of $P$ is queried at most once and cached, and that subsequent queries are retrieved from the cache.

Let $\pi_M$ be a self-delimiting program for $M$. Let $p_n \in \{0,1\}^{\#(S\upharpoonright n, M^P)}$ be the oracle bits of $P$ queried by $M$ on input $n$, in the order in which they are queried. Recall the self-delimiting encoding function enc. For each $n \in \mathbb{N}$, let $\pi_n = \pi_{M'}\pi_M\mathrm{enc}(n)\mathrm{enc}(p_n)$, where $\pi_{M'}$ is a self-delimiting program that simulates $M$, encoded by $\pi_M$, on input $n$, encoded by $\mathrm{enc}(n)$, with oracle $P$, encoded by $\mathrm{enc}(p_n)$. When $M$ makes its $i^{\text{th}}$ query to a bit of $P$, the bit $p_n[i]$ is returned. Since $M$ queries each bit of $P$ at most once, the bit from $p_n$ will be correct, no matter what index was queried by $M$, since the bits of $p_n$ are arranged in the order in which $M$ makes its queries.

Then $U(\pi_n) = S \upharpoonright n$, so if there exists $s \in \mathrm{bound}(\Delta)$ such that $M$ uses at most $s(n)$ space on input $n$, there exists $q \in \mathrm{bound}(\Delta)$ such that, for all $n \in \mathbb{N}$, $\mathrm{KS}^q(S \upharpoonright n) \leq |\pi_n|$.

By Theorem 3.10,

$$
\begin{aligned}
& \dim_\Delta(S) \\
= & \inf_{q \in \mathrm{bound}(\Delta)} \liminf_{n\to\infty} \frac{\mathrm{KS}^q(S \upharpoonright n)}{n} \\
\leq & \liminf_{n\to\infty} \frac{|\pi_{M'}\pi_M \mathrm{enc}(n)\mathrm{enc}(p_n)|}{n} \\
\leq & \liminf_{n\to\infty} \frac{|\pi_{M'}\pi_M| + \log n + 2\log\log n + \#(S\upharpoonright n, M^P) + 2\log\#(S\upharpoonright n, M^P) + 6}{n} \\
= & \liminf_{n\to\infty} \frac{\#(S\upharpoonright n, M^P)}{n} \\
= & \rho^-_M(S, P),
\end{aligned}
$$

whence $\dim(S) \leq \rho^-_\Delta(S)$. Similarly, $\mathrm{Dim}(S) \leq \rho^+_\Delta(S)$. □

The following theorem is used to construct Δ-dimension extractors, and to give new characterizations of these dimensions for $\Delta = \mathrm{comp}$ or $\mathrm{p}_i\mathrm{space}$. The theorem also holds for $\mathrm{p}_i$ dimension, but the second application, the characterization of dimension in terms of optimal sequence decompression, seems to require a $\mathrm{p}_i$-time-bounded Kolmogorov complexity characterization of $\mathrm{p}_i$ dimension, which at the present time is unknown. Furthermore, the theorem only holds for polynomial time and above, and does not include linear time.

**Theorem 4.5.** *Let $i \in \mathbb{N}$ and $\Delta \in \{\mathrm{comp}, \mathrm{p}_i\mathrm{space}, \mathrm{p}_{i+1}\}$. For all $S \in \mathbf{C}$ and $\delta > 0$, there is a sequence $P \in \mathbf{C}$ and an OTM $M$ such that*

1. $S \equiv^\Delta_\mathrm{T} P$, *with* $S \leq^\Delta_\mathrm{T} P$ *via* $M$.
2. $\rho^-_M(S, P) \leq \dim_\Delta(S) + \delta$.
3. $\rho^+_M(S, P) \leq \mathrm{Dim}_\Delta(S) + \delta$.

*Proof idea:* If the Δ-dimension of $S$ is small, then a Δ-computable martingale $d$ performs well on $S$. Thus, if we have already computed a prefix $S \upharpoonright n$ of $S$, then *on average*, $d$ increases its capital more on the next $k$ bits of $S$ than it would on other $k$-bit strings that could extend $S \upharpoonright n$. This places the next $k$ bits of $S$ in a small (on average) subset of $\{0,1\}^k$, namely, those strings on which $d$ increases its capital above a certain rational threshold $c$, which is chosen to be slightly smaller than $d(S \upharpoonright (n+k))$, the amount of capital made after the next $k$ bits of $S$. Since $d$ is Δ-computable, it is possible to enumerate strings from this small set by evaluating $d$ in parallel on all possible length-$k$ extensions of $S \upharpoonright n$, and outputting a string $u \in \{0,1\}^k$ whenever $d((S \upharpoonright n)u)$ is greater than $c$. We will encode the next $k$ bits of $S$ as an index into this set, where the index will represent the order in which this parallel evaluation enumerates the string we want – the next $k$ bits of $S$. This technique is similar to that used by Merkle and Mihailović [37] to prove Theorem 5.4.

We require two lemmas to prove Theorem 4.5. Lemma 4.7 shows that the average number of bits needed to encode the index of a length-$k$ extension of $S \upharpoonright n$ is close to the dimension of $S$ times $k$. We will also need to encode the threshold $c$ into the oracle sequence. Lemma 4.6 shows that we can find a rational threshold $c$ that requires so few bits to represent that it will not affect the compression ratio when added to the oracle sequence, yet which is still a close enough approximation to $d(S \upharpoonright (n+k))$ to keep the index length of Lemma 4.7 small.

In the following lemma, intuitively, the values $k_i$, $n_i$, $r_i$, and $c_i$ respectively represent the number of bits in the $i^{\text{th}}$ block of the sequence $S$, the number of bits in the first $i$ blocks, the value of a martingale after reading the first $n_i$ bits, and a rational approximation to that value from below.

**Lemma 4.6.** *There exists a constant $C$ such that the following holds. Let $N \in \mathbb{Z}^+$. Let $k_0 = n_0 = 2$. For all $i \in \mathbb{Z}^+$, let $k_i = \lceil N \log n_{i-1} \rceil$ and $n_i = n_{i-1} + k_i$. Let $r_i \in [1, 2^{n_i}] \cap \mathbb{Q}_2$. Then for all but finitely many $i \in \mathbb{N}$, there exists $c_i \in \mathbb{Q}_2$ such that $r_i \left(1 - \frac{1}{k_i^2}\right) \leq c_i < r_i$ and $\mathrm{K}^{Cn_i}(c_i) \leq k_i \cdot 2/N$. Furthermore, such a program can be computed from $k_i$ and $r_i$ in at most $Cn_i$ steps.*

*Proof.* We prove the cases $k_i^2 \leq r_i \leq 2^{n_i+1}$ and $1 \leq r_i < k_i^2$ separately.

Suppose that $k_i^2 \leq r_i \leq 2^{n_i+1}$. In this case we will choose $c_i$ to be an integer. Set $m \in \mathbb{Z}^+$ such that $2^{m-1} < k_i^2 \leq 2^m$. Since $r_i \geq k_i^2 > 2^{m-1}$, $\lceil \log r_i \rceil > m - 1$.

If $r_i$ is an integer ending in at least $\lceil \log r_i \rceil - m$ bits, let $c_i = r_i - 1$. Otherwise, let $c_i \in \mathbb{Z}^+$ be the integer whose binary representation is $x0^{\lceil \log r_i \rceil - m}$, where $x \in \{0,1\}^m$ is the first $m$ bits of $\lfloor r_i \rfloor$. Since $c_i$ shares its first $m$ bits with $r_i$ or is equal to $r_i - 1$,

$$r_i - c_i \leq 2^{\lceil \log r_i \rceil - m} - 1 \leq \frac{r_i + 2}{2^m} - 1 \leq \frac{r_i}{k_i^2},$$

so $r_i \left(1 - \frac{1}{k_i^2}\right) \leq c_i < r_i$. $c_i$ can be fully described by the first $m$ bits of $r_i$, along with the binary representation of the number $\lceil \log r_i \rceil - m$ of 0's that follow. Thus, describing $c_i$ requires no more than

$$\begin{aligned} m + \log(\lceil \log r_i \rceil - m) &\leq \log k_i^2 + \log n_i \\ &\leq \log k_i^2 + \log 2^{k_{i+1}/N} \\ &\leq 2k_i/N \end{aligned}$$

bits, for all but finitely many $i \in \mathbb{N}$.

Now suppose that $1 \leq r_i < k_i^2$. We let $c_i$ approximate $r_i$ by the binary integer $\lfloor r_i \rfloor$, plus a finite prefix of the bits to the right of $r_i$'s decimal point in binary form. If $x.S$ is the binary representation of $r_i$, where $x \in \{0,1\}^*$ and $S \in \mathbf{C}$, let $c_i \in \mathbb{Q}_2^+$ be represented by $x.y$, where $y \sqsubseteq S$. Then $r_i - c_i \leq 2^{-|y|}$.

Since $r_i < k_i^2$, $|x| \leq \log k_i^2 = 2 \log k_i$. We need $r_i - c_i \leq r_i/k_i^2$. Since $r_i - c_i \leq 2^{-|y|}$, it suffices to choose $y \sqsubseteq S$ such that $2^{-|y|} \leq r_i/k_i^2$, or $|y| \geq \log(k_i^2/r_i)$. Let $|y| = \lceil \log(k_i^2/r_i) \rceil \leq 2 \log k_i$, since $r_i \geq 1$. Thus $|x| + |y| \leq 4 \log k_i$, so describing $c_i$ requires at most $4 \log k_i$ bits.

If this results in $c_i = r_i$, rather than $c_i < r_i$, then $r_i$ is already a dyadic rational requiring no more than $|y|$ bits to the right of the decimal point. In this case, let $c_i = r_i - 2^{-|y|} \geq r_i \left(1 - \frac{1}{k_i^2}\right)$ instead.

Let $\pi(c_i)$ be one of the two programs just described for computing $c_i$ from $r_i$. It is clear that $\pi(c_i)$ runs in $O(n_i)$ time. We now demonstrate that $\pi(c_i)$ itself can be computed from $i$ and $r_i$ in $O(n_i)$ time. $\pi(c_i)$ is simply constructed from initial bits of $r_i$ in either case. It follows that in the first case, $\pi(c_i)$ can be created in $O(n_i)$ time, since $\log r_i \leq n_i$, and the first case uses only the integral part of $r_i$. In the second case, we have already shown that the integral part $x$ of $c_i$ and the fractional part $y$ of $c_i$ each consist of $O(\log k_i)$ bits of $r_i$. Therefore, in the second case, $\pi(c_i)$ can be created in $O(n_i)$ time. □

For $t \in \mathbb{R}$, $c \in \mathbb{Q}$, $s \in \{0,1\}^*$, $k \in \mathbb{N}$, and $d : \{0,1\}^* \to [0, \infty)$ a $t$-gale, define

$$A^{(k)}_{d,c,s} = \left\{ \ u \in \{0,1\}^k \ \middle| \ d(su) > c \ \right\} \tag{4.7}$$

to be the set of all length-$k$ extensions of $s$ on which $d$ makes at least $c$ capital. The following lemma shows that $|A^{(k)}_{d,c,s}|$ is small on average if $d$ makes a lot of capital on a sequence beginning with prefix $s$, if $c$ is close to the capital that $d$ has after reading $k$ bits beyond $s$.

**Lemma 4.7.** *Let $S \in \mathbf{C}$, $r \geq t > 0$, and let $d$ be a martingale such that $d^{(t)}$ succeeds on $S$ and $d^{(r)}$ strongly succeeds on $S$. Write $S = \widehat{s}_0\widehat{s}_1\widehat{s}_2\ldots$, where, for all $i \in \mathbb{N}$, $k_i = |\widehat{s}_i|$, and $n_i = |\widehat{s}_0 \ldots \widehat{s}_i|$. Let $c_i \in \mathbb{R}$ satisfy $d(S \upharpoonright n_i)g(i) \leq c_i \leq d(S \upharpoonright n_i)$, where $g(i) \in (0,1)$ satisfies $-\sum_{j=2}^{i} \log g(j) = o(n_i)$ as $i \to \infty$. Then*

$$\limsup_{i \to \infty} \frac{\sum_{j=0}^{i} \log \left| A^{(k_j)}_{d,c_j,S \upharpoonright n_{j-1}} \right|}{n_i} \leq r, \tag{4.8}$$

*and, if $k_i = o(n_i)$ as $i \to \infty$, then*

$$\liminf_{i \to \infty} \frac{\sum_{j=0}^{i} \log \left| A^{(k_j)}_{d,c_j,S \upharpoonright n_{j-1}} \right|}{n_i} \leq t, \tag{4.9}$$

*Proof.* We first show that (4.9) holds. Let $t' > t$, and, for all $i \in \mathbb{N}$, let $A_i = A^{(k_i)}_{d,c_i,S \upharpoonright n_{i-1}}$. It suffices to show that, for infinitely many $i \in \mathbb{N}$, $\sum_{j=0}^{i} \log |A_j| \leq t' n_i$. Since $d^{(t)}$ succeeds on $S$, for infinitely many $n \in \mathbb{N}$,

$$d(S \upharpoonright n) \geq 2^{(1-t)n} d(\lambda). \tag{4.10}$$

A martingale can at most double its capital after every bit, and each index $n$ with $n_i \leq n < n_{i+1}$ is at most $k_i$ bits beyond $n_i$. It follows that for infinitely many $i \in \mathbb{N}$,

$$d(S \upharpoonright n_i) \geq 2^{(1-t)n_i - k_i} d(\lambda). \tag{4.11}$$

For all $i \in \mathbb{N}$, set $l_i \in \mathbb{R}$ such that $d(S \restriction n_i) = 2^{k_i - l_i} d(S \restriction n_{i-1})$. By induction on $i$,

$$d(S \restriction n_i) = d(\lambda) \prod_{j=0}^{i} 2^{k_j - l_j}. \tag{4.12}$$

Then, by equations (4.11) and (4.12), and the fact that $\sum_{j=0}^{i} k_i = n_i$, for infinitely many $i \in \mathbb{N}$,

$$\begin{aligned} \prod_{j=0}^{i} 2^{k_j - l_j} \geq 2^{(1-t)n_i - k_i} \quad &\Longrightarrow \quad \sum_{j=0}^{i} (k_j - l_j) \geq (1-t)n_i - k_i \\ &\Longrightarrow \quad \sum_{j=0}^{i} l_j \leq t n_i + k_i. \end{aligned}$$

Recall that $c_i \geq d(S \restriction n_i) g(i) = g(i) 2^{k_i - l_i} d(S \restriction n_{i-1})$. By Corollary 3.3 (take $k = k_i, l = l_i, \alpha = 1 - \frac{1}{i^2}, w = S \restriction n_{i-1}$) and the definition of $l_i$, it follows that $|A_i| \leq 2^{l_i}/g(i)$, and so $\log |A_i| \leq l_i - \log g(i)$. Let $c_{0,1} = \log |A_0| + \log |A_1| - l_0 - l_1$. Then

$$\begin{aligned} \sum_{j=0}^{i} \log |A_j| &\leq \sum_{j=0}^{i} l_j - \sum_{j=2}^{i} \log g(i) + c_{0,1} \\ &\leq t n_i + k_i - \sum_{j=2}^{i} \log g(i) + c_{0,1} \\ &= t' n_i + (t - t') n_i + k_i - \sum_{j=2}^{i} \log g(i) + c_{0,1}. \end{aligned} \tag{4.13}$$

$t < t'$, $k_i = o(n_i)$, and $\sum_{j=2}^{i} \log g(i) = o(n_i)$, so for infinitely many $i$, $\sum_{j=0}^{i} \log |A_j| \leq t' n_i$.

The proof of (4.8) is similar, replacing "for infinitely many $i$" conditions with "for all but finitely many $i$." The only difference is that (4.10) holds for all but finitely many $n$, and so there is no need to derive (4.11). Consequently, the term $k_i$ does not appear on the right-hand side of (4.13), and so the condition $k_i = o(n_i)$ is not necessary to show that (4.8) holds. □

Let $t \in \mathbb{R}$, $c \in \mathbb{Q}$, $s \in \{0,1\}^*$, $k \in \mathbb{N}$, and let $d : \{0,1\}^* \to [0, \infty)$ be a $t$-gale. If $d$ is exactly $\Delta$-computable and $u \in A^{(k)}_{d,c,s}$, then the following procedure computes the index of $u$ in a lexicographical ordering of $A^{(k)}_{d,c,s}$.

$\mathrm{ind}^{(k)}_{d,c,s} \left( u \in \{0,1\}^k \right)$

$i' \leftarrow 0$

**for** each $u' \in \{0,1\}^k$ in lexicographical order

    **do if** $d(su') > c$

        **then if** $u' = u$

            **then** output $i'$ and exit

            **else** $i' \leftarrow i' + 1$

If $u \notin A^{(k)}_{d,c,s}$, $\mathrm{ind}^{(k)}_{d,c,s}(u)$ is undefined. For all $u \in \{0,1\}^k$, $\mathrm{ind}^{(k)}_{d,c,s}(u) \leq \left|A^{(k)}_{d,c,s}\right|$ when it is defined. The computation of $\mathrm{str}^{(k)}_{d,c,s} : \mathbb{N} \to \{0,1\}^k$, the inverse of $\mathrm{ind}^{(k)}_{d,c,s}$, is similar:

$\mathrm{str}^{(k)}_{d,c,s}\,(i \in \mathbb{N})$
$i' \leftarrow 0$
**for** each $u' \in \{0,1\}^k$ in lexicographical order
    **do if** $d(su') > c$
        **then if** $i' = i$
            **then** output $u'$ and exit
            **else** $i' \leftarrow i' + 1$

Both $\mathrm{ind}^{(k)}_{d,c,s}$ and $\mathrm{str}^{(k)}_{d,c,s}$ are *uniformly* $\Delta$*-computable* for all $d$, $c$, $s$, and $k$, in the sense that each may be implemented by a single $\Delta$-bounded Turing machine taking $d$, $c$, $s$, and $k$ as auxiliary input, provided $d$ is exactly $\Delta$-computable.

*Proof of Theorem 4.5.* If $\dim_\Delta(S) = 1$, then the trivial "bit-copier" OTM $M$ suffices to compute $P = S$, where $\rho^-_M(S,P) = \rho^+_M(S,P) = \dim_\Delta(S) = \mathrm{Dim}_\Delta(S) = 1$, so assume that $\dim_\Delta(S) < 1$.

Let $t, r \in \mathbb{Q}_2$ such that $\dim_\Delta(S) < t < \dim_\Delta(S) + \delta$ and $\mathrm{Dim}_\Delta(S) < r < \mathrm{Dim}_\Delta(S) + \delta$. Then by Observation 3.11, there is a $\Delta$-computable martingale $d$ such that $d^{(t)}$ succeeds on $S$ and $d^{(r)}$ strongly succeeds on $S$. By Observation 3.5, $\dim_\Delta(S) < 1$ implies that we may assume that $d$ strongly succeeds on $S$. By Lemma 3.4, we may assume that $d$ is dyadically $\Delta$-computable. By Observation 3.12, we may assume that $d(\lambda) = 1$.

Let $N = \left\lceil \max\left\{ \frac{2}{\dim_\Delta(S) + \delta - r}, \frac{2}{\mathrm{Dim}_\Delta(S) + \delta - t} \right\} \right\rceil$. Let $k_0 = n_0 = 2$. For all $i \in \mathbb{Z}^+$, let $k_i = \lceil N \log n_{i-1} \rceil$ and $n_i = n_{i-1} + k_i$. Write $S = \widehat{s}_0 \widehat{s}_1 \widehat{s}_2 \ldots$, where, for all $i \in \mathbb{N}$, $k_i = |\widehat{s}_i|$, and $n_i = |\widehat{s}_0 \ldots \widehat{s}_i|$. By Observation 3.1, $d(S \upharpoonright n_i) \leq 2^{n_i}$ for all $i \in \mathbb{N}$. Since $d$ strongly succeeds on $S$, for all but finitely many $i$, $d(S \upharpoonright n_i) \geq 1$. Let $C \in \mathbb{N}$ be as in Lemma 4.6. For all such $i \in \mathbb{N}$, taking $r_i = d(S \upharpoonright n_i)$, choose $c_i \in \mathbb{Q}^+$ for $i$, $a$, and $r_i$ as in Lemma 4.6, and let $\pi(c_i)$ represent a program testifying that $\mathrm{K}^{Cn_i}(c_i) \leq k_i \cdot 2/N$, which can be computed from $i$, $d(\lambda)$, and $d(S \upharpoonright n_i)$ in at most $Cn_i$ steps.

Let $P = p_0 p_1 p_2 \ldots$, where, for all $i \in \mathbb{N}$,

$$p_i = \mathrm{enc}\left(\mathrm{ind}^{(k_i)}_{d,c_i,S\upharpoonright n_{i-1}}(\widehat{s}_i)\right) \pi(c_i).$$

Because $\mathrm{str}^{(k_i)}_{d,c_i,S\upharpoonright n_{i-1}}$ is an inverse of $\mathrm{ind}^{(k_i)}_{d,c_i,S\upharpoonright n_{i-1}}$, we can write each $\widehat{s}_i$ as

$$\widehat{s}_i = \mathrm{str}^{(k_i)}_{d,c_i,S\upharpoonright n_{i-1}}\left(\mathrm{ind}^{(k_i)}_{d,c_i,S\upharpoonright n_{i-1}}(\widehat{s}_i)\right).$$

Since $\mathrm{ind}^{(k_i)}_{d,c_i,S\upharpoonright n_{i-1}}$, $\mathrm{str}^{(k_i)}_{d,c_i,S\upharpoonright n_{i-1}}$ and $d$ are all $\Delta$-computable, $P \equiv^\Delta_\mathrm{T} S$. Note in particular that, since $k_i = O(\log n_i)$, the search over all strings of length $k_i$ done in the computation of $\mathrm{ind}^{(k_i)}_{d,c_i,S\upharpoonright n_{i-1}}$ and $\mathrm{str}^{(k_i)}_{d,c_i,S\upharpoonright n_{i-1}}$ requires searching only a polynomial (in $n_i$) number of strings.

Let $M$ be the OTM such that $M^P$ computes $S$. It suffices to show that $\rho^-_M(S,P) \leq \dim_\Delta(S) + \delta$ and $\rho^+_M(S,P) \leq \mathrm{Dim}_\Delta(S) + \delta$. Recall that $k_i = n_i - n_{i-1}$, and $k_i$ grows unboundedly with $i$, so $i = o(n_i)$ as $i \to \infty$. Then

$$\begin{aligned} -\sum_{j=2}^{i} \log\left(1 - \frac{1}{k_j^2}\right) &= \sum_{j=2}^{i} [2\log k_j - \log(k_j+1) - \log(k_j - 1)] \\ &\leq 2\sum_{j=2}^{i} [\log k_j - \log(k_j - 1)] \\ &\leq 2\sum_{j=2}^{i} [\log(N\log(n_{j-1}) + 1) - \log(N\log(n_{j-1}) - 1)] \\ &= O(i) = o(n_i), \end{aligned}$$

since $\lim_{j\to\infty} [\log(N\log(n_{j-1}) + 1) - \log(N\log(n_{j-1}) - 1)] = 0$. Thus $g(j) = 1 - \frac{1}{k_j^2}$ satisfies the requirement $-\sum_{j=2}^{i} \log g(j) = o(n_i)$ in the hypothesis of Lemma 4.7. Then

$$\begin{aligned} \rho^+_M(S,P) &= \limsup_{n\to\infty} \frac{\#(S\upharpoonright n, M^P)}{n} = \limsup_{i\to\infty} \frac{\#(S\upharpoonright n_i, M^P)}{n_i} \qquad \text{since } k_i = o(n_i) \\ &= \limsup_{i\to\infty} \frac{\sum_{j=0}^{i} \left|\mathrm{enc}\left(\mathrm{ind}^{(k_j)}_{d,c_j,S\upharpoonright n_{j-1}}(\widehat{s}_j)\right)\pi(c_j)\right|}{n_i} \\ &= \limsup_{i\to\infty} \left( \frac{\sum_{j=0}^{i} |\pi(c_j)|}{n_i} + \frac{\sum_{j=0}^{i} \log \mathrm{ind}^{(k_j)}_{d,c_j,S\upharpoonright n_{j-1}}(\widehat{s}_j)}{n_i} \right) \\ &\leq \limsup_{i\to\infty} \left( \frac{\sum_{j=0}^{i} k_j \cdot 2/N}{\sum_{j=0}^{i} k_j} + \frac{\sum_{j=0}^{i} \log \mathrm{ind}^{(k_j)}_{d,c_j,S\upharpoonright n_{j-1}}(\widehat{s}_j)}{n_i} \right) \\ &\leq 2/N + r \leq \dim_\Delta(S) + \delta \qquad \text{by Lemma 4.7.} \end{aligned}$$

Similarly, $\rho^-_M(S,P) \leq \mathrm{Dim}_\Delta(S) + \delta$. □

We now arrive at the main result of this section. It is a $\Delta$-bounded extension and improvement of Corollary 4.3. It states that $\Delta$-strong dimension may be extracted, using $\Delta$-Turing reductions, from any sequence of positive $\Delta$-dimension, for $\Delta = \mathrm{comp}$, $\mathrm{p}_i\mathrm{space}$ (for $i \geq 0$), or $\mathrm{p}_i$ (for $i \geq 1$). Furthermore, Theorem 4.8 achieves a partial extraction of dimension in addition to strong dimension. The reduction is not uniform: the OTMs used depend on the sequence from which dimension is being extracted. The case of Theorem 4.8 for $\Delta = \mathrm{comp}$ was independently discovered by Laurent Bienvenu [6]. Theorem 4.8 has been extended to constructive dimension by Bienvenu, Doty, and Stephan [7].

**Theorem 4.8.** *Let $i \in \mathbb{N}$ and $\Delta \in \{\mathrm{comp}, \mathrm{p}_i\mathrm{space}, \mathrm{p}_{i+1}\}$. For each $\epsilon > 0$ and each $S \in \mathbf{C}$ such that $\dim_\Delta(S) > 0$, there exists $P \in \mathbf{C}$ such that $P \equiv^\Delta_\mathrm{T} S$, $\dim_\Delta(P) \geq \frac{\dim_\Delta(S)}{\mathrm{Dim}_\Delta(S)} - \epsilon$, and $\mathrm{Dim}_\Delta(P) \geq 1 - \epsilon$.*

*Proof.* Let $i \in \mathbb{N}$ and $\Delta \in \{\mathrm{comp}, \mathrm{p}_i\mathrm{space}\}$. We first prove this case, and at the end of the proof we handle the case $\Delta = \mathrm{p}_i$ for $i \in \mathbb{Z}^+$.

Let $S$ and $\epsilon$ be as in the statement of the theorem. Let $0 < \delta < \epsilon \cdot \dim_\Delta(S)/2$. Choose $P \in \mathbf{C}$, $M_d \in \mathrm{OTM}$, and $M_e \in \mathrm{OTM}$ for $S$ and $\delta$ as in Theorem 4.5 such that $P \equiv_{\mathrm{T}}^{\Delta} S$ and, letting $n_p = \#(S \upharpoonright n, M_d^P)$ for all $n \in \mathbb{N}$,

$$
\begin{aligned}
& \liminf_{n\to\infty} \frac{n_p}{n} < \dim_\Delta(S) + \delta \\
\implies & \inf_{q \in \mathrm{bound}(\Delta)} \liminf_{m\to\infty} \frac{\mathrm{KS}^q(S \upharpoonright m)}{m} > \liminf_{n\to\infty} \frac{n_p}{n} - \delta \qquad \text{by Theorem 3.10} \\
\implies & (\forall q \in \mathrm{bound}(\Delta)) \ \liminf_{m\to\infty} \frac{\mathrm{KS}^q(S \upharpoonright m)}{m} > \liminf_{n\to\infty} \frac{n_p}{n} - \delta \\
\implies & (\forall q \in \mathrm{bound}(\Delta))(\exists^\infty n \in \mathbb{N}) \ \liminf_{m\to\infty} \frac{\mathrm{KS}^q(S \upharpoonright m)}{m} > \frac{n_p}{n} - \delta \\
\implies & (\forall q \in \mathrm{bound}(\Delta))(\exists^\infty n \in \mathbb{N}) \ \mathrm{KS}^q(S \upharpoonright n) > n_p - \delta n. \qquad (4.14)
\end{aligned}
$$

Since $\delta < \epsilon \cdot \dim_\Delta(S)/2$, by Theorem 3.10,

$$(\forall q \in \mathrm{bound}(\Delta))(\forall^\infty n \in \mathbb{N}) \ \mathrm{KS}^q(S \upharpoonright n) > \frac{2}{\epsilon}\delta n. \qquad (4.15)$$

Note that the construction of $M_d$ in the proof of Theorem 4.5 is such that entire prefixes of the oracle sequence are queried at once: whenever the bit at index $i \in \mathbb{N}$ is queried, all bits $j < i$ are also queried. Thus, a program $M'$ simulating $M_d$ with the first $n_p$ bits of $P$ can calculate $S \upharpoonright n$, and $M'$ can be encoded in $|\mathrm{enc}(P \upharpoonright n_p)| + O(1)$ bits. Thus there is a constant $c$ and, since $M_d$ uses $\Delta$ space, there exists $t \in \mathrm{bound}(\Delta)$ such that $\mathrm{KS}^t(S \upharpoonright n) \leq n_p + 2\log n_p + c$, which together with (4.15) implies that

$$(\forall^\infty n \in \mathbb{N}) \ \delta n < \frac{\epsilon}{2}(n_p + 2\log n_p + c). \qquad (4.16)$$

Combining (4.14) and (4.16),

$$(\forall q \in \mathrm{bound}(\Delta))(\exists^\infty n \in \mathbb{N}) \ \mathrm{KS}^q(S \upharpoonright n) > n_p - \frac{\epsilon}{2}\left(p_n + 2\log n_p + c\right). \qquad (4.17)$$

Suppose for the sake of contradiction that $\mathrm{Dim}_\Delta(P) < 1 - \epsilon$. Then by Theorem 3.10, it would be the case that

$$(\exists s \in \mathrm{bound}(\Delta))(\forall^\infty m \in \mathbb{N}) \ \mathrm{KS}^s(P \upharpoonright m) < m - \epsilon m. \qquad (4.18)$$

Since $\dim_\Delta(S) > 0$, $S$ is not $\Delta$-computable (see [33], Lemma 4.13), and since $S$ is $\Delta$-computable by $M_d$ with access to $P$, $n_p$ must grow unboundedly with $n$. A program that produces $P \upharpoonright n_p$ in $s(n_p)$ space can be used in conjunction with $M_d$ (which uses at most $t(n)$ space) to produce $S \upharpoonright n$ in at most $t(n) + s(n_p)$ space. For the case $\Delta = \mathrm{comp}$, $t(n) + s(n_p)$ is bounded by a computable function of $n$. By part 3 of Theorem 4.5, $n_p = O(n)$, so, for the case $\Delta = \mathrm{p}_i\mathrm{space}$, the space bound $n \mapsto t(n) + s(n_p)$ is contained in $G_i = \mathrm{bound}(\mathrm{p}_i\mathrm{space})$. Then for a suitable constant $c' \approx |M_d|$,

$$\begin{aligned}(\exists q \in \mathrm{bound}(\Delta))(\forall^\infty n \in \mathbb{N})\ \mathrm{KS}^q(S \upharpoonright n) &\leq \mathrm{KS}^s(P \upharpoonright n_p) + c' \\ &< n_p - \epsilon n_p + c' \\ &< n_p - \frac{\epsilon}{2}\left(p_n + 2\log n_p + c\right). \end{aligned} \tag{4.19}$$

But (4.19) contradicts (4.17). Hence, $\mathrm{Dim}_\Delta(P) \geq 1 - \epsilon$.

To see that $\dim_\Delta(P) \geq \frac{\dim_\Delta(S)}{\mathrm{Dim}_\Delta(S)} - \epsilon$, let $0 < d < \dim_\Delta(S)$, $D' > D > \mathrm{Dim}_\Delta(S)$, and $0 < \delta < \min\{\dim_\Delta(S) - d, D - \mathrm{Dim}_\Delta(S)\}$. (Note that our previous requirement that $0 < \delta < \epsilon \cdot \dim_\Delta(S)/2$ does not contradict this.)

Let $q \in \mathrm{bound}(\Delta)$. It suffices to prove that, for all but finitely many $n$, $\mathrm{KS}^{r+q}(P \upharpoonright n) \geq \frac{d}{D'}n$, where $r \in \mathrm{bound}(\Delta)$ is given below.

By Theorem 4.5, $P \equiv^\Delta_\mathrm{T} S$ and $\rho^+_{M_d}(S, P) < D$. Recall that for each $n \in \mathbb{N}$, $n_p = \#(S \upharpoonright n, M^P_d)$. Then there exists $r \in \mathrm{bound}(\Delta)$ such that, for all but finitely many $n$,

$$dn < \mathrm{KS}^q(S \upharpoonright n) < \mathrm{KS}^{r+q}(P \upharpoonright n_p) + 2\log n. \tag{4.20}$$

The first inequality holds because $d < \dim_\Delta(S)$, and the second because $M_d^{P\upharpoonright n_p}$ together with a binary encoding of $n$ (requiring less than $2\log n$ bits) is a program computable in $r(n) + q(n)$ space to output $S \upharpoonright n$.

By the fact that $\rho^+_{M_d}(S, P) < D$, the assumption that $D' > D$, and (4.20), for all but finitely many $n$,

$$\begin{aligned} Dn > n_p &\implies D'n - 2\frac{D'}{d}\log n > n_p \\ &\implies dn - 2\log n > \frac{d}{D'}n_p \\ &\implies \mathrm{KS}^{r+q}(P \upharpoonright n_p) > \frac{d}{D'}n_p. \end{aligned} \tag{4.21}$$

Recall from the proof of Theorem 4.5 that $|\widehat{s}_i|$, the length of the $i^{\text{th}}$ block in $S$, is $o(i)$, and that $|p_i|$, the length of the $i^{\text{th}}$ block of $P$, is at most $|\widehat{s}_i| + o(|\widehat{s}_i|)$. This implies that $|p_i| = o(n_p)$ if $p_i$ occurs on prefix $n_p$. Thus, for all $n$, $|n_p - (n+1)_p| = o(n_p)$. Therefore, (4.21) implies that, for all but finitely many $n \in \mathbb{N}$, $\mathrm{KS}^{r+q}(P \upharpoonright n) \geq \frac{d}{D'}n$. This proves the theorem for $\Delta \in \{\mathrm{comp}, \mathrm{p}_i\mathrm{space}\}$.

We next show that the theorem holds for $\Delta = \mathrm{p}_i$, where $i \in \mathbb{Z}^+$. Note that the OTMs used in the proof of Theorem 4.5 compute $P$ from $S$ (respectively, $S$ from $P$) using $S \upharpoonright n$ to compute $P \upharpoonright n_p$ (respectively, $P \upharpoonright n_p$ to compute $S \upharpoonright (n-1)$), where, for all $n \in \mathbb{N}$,

$$(n+1)_p - n_p = O(\log n), \tag{4.22}$$

because the blocks in which $S$ and $P$ are processed have length logarithmic in the length of the prefixes they extend. This implies that there exist $\mathrm{p}_i$-computable functions $C$ :

$\{0,1\}^* \to \{0,1\}^*$ and $D : \{0,1\}^* \times \mathbb{N} \to \{0,1\}^*$ such that, for all $n \in \mathbb{N}$, $D(C(S \upharpoonright n), n) = S \upharpoonright n$; $C$ simply encodes however many whole blocks of $S$ fit into $S \upharpoonright n$ as the corresponding blocks from $P$, and the last partial block of $S$ may simply be copied bit-for-bit. Equation (4.22) tells us that the last block's length is negligible, which implies that $(C, D)$ achieves a compression ratio asymptotically as good as that of Theorem 4.5. Therefore

$$\liminf_{n\to\infty} \frac{|C(S \upharpoonright n)|}{n} \le \dim_{\mathrm{p}_i}(S) + \delta \tag{4.23}$$

and

$$\limsup_{n\to\infty} \frac{|C(S \upharpoonright n)|}{n} \le \mathrm{Dim}_{\mathrm{p}_i}(S) + \delta. \tag{4.24}$$

Equation (4.22) also implies that $(C, D)$ satisfies (3.1) and therefore, $(C, D)$ is a $\mathrm{p}_i$-compressor that does not start from scratch. Although Lemma 4.4 is not known to hold for $\mathrm{p}_i$-dimension, Theorem 3.14 and inequalities (4.23) and (4.24) tell us that $(C, D)$ achieves a compression ratio within $\delta$ of the optimal compression ratio achievable by $\mathrm{p}_i$-compressors that do not start from scratch. Hence, by an incompressibility argument similar to that given in the first part of the proof (and similar to the proof of Theorem 4.14), $\dim_{\mathrm{p}_i}(P) \ge \frac{\dim_{\mathrm{p}_i}(S)}{\mathrm{Dim}_{\mathrm{p}_i}(S)} - \epsilon$, and $\mathrm{Dim}_{\mathrm{p}_i}(P) \ge 1 - \epsilon$. □

**Corollary 4.9.** *Let $i \in \mathbb{N}$ and $\Delta \in \{\mathrm{comp}, \mathrm{p}_i\mathrm{space}, \mathrm{p}_{i+1}\}$. For each $\epsilon > 0$ and each $\Delta$-regular $S \in \mathbf{C}$ such that $\dim_\Delta(S) > 0$, there exists $P \in \mathbf{C}$ such that $P \equiv_{\mathrm{T}}^{\Delta} S$ and $\dim_\Delta(P) \ge 1 - \epsilon$.*

Fortnow, Hitchcock, Pavan, Vinodchandran, and Wang [17] have shown a similar extraction result for constructive, computable, and space-bounded strong dimension, which starts from the weaker hypothesis that only the *strong* dimension of the input sequence is positive. Furthermore, the extractor runs in polynomial time. Lemma 4.2 of [17] requires advice and is stated only for pspace dimension, but can be improved to eliminate the advice [25]. The technique also works for any dimension having a space-bounded Kolmogorov complexity characterization, leading to the following theorem.

**Theorem 4.10** ( [17]). *Let $i \in \mathbb{N}$ and $\Delta \in \{\mathrm{comp}, \mathrm{p}_i\mathrm{space}\}$. For each $\epsilon > 0$ and $S \in \mathbf{C}$,*

1. *If $\mathrm{Dim}(S) > 0$, then there exists $P \in \mathbf{C}$ such that $P \equiv_{\mathrm{T}}^{\mathrm{p}} S$ and $\mathrm{Dim}(P) \ge 1 - \epsilon$.*
2. *If $\mathrm{Dim}_\Delta(S) > 0$, then there exists $P \in \mathbf{C}$ such that $P \equiv_{\mathrm{T}}^{\mathrm{p}} S$ and $\mathrm{Dim}_\Delta(P) \ge 1 - \epsilon$.*

### 4.2.2 Finite-State Dimension Extractors

We next show that information lossless finite-state transducers can extract finite-state dimension [12]. Finite-state dimension has multiple equivalent definitions [9, 12, 14, 23]. For this paper, we exclusively use the characterization given in [12] in terms of information lossless finite-state transducers.

A *finite-state transducer (FST)* is a 4-tuple $T = (Q, \delta, \nu, q_0)$, where

- $Q$ is a nonempty, finite set of *states*,
- $\delta : Q \times \{0,1\} \to Q$ is the *transition function*,
- $\nu : Q \times \{0,1\} \to \{0,1\}^*$ is the *output function*,
- $q_0 \in Q$ is the *initial state*.

For all $x \in \{0,1\}^*$ and $b \in \{0,1\}$, define the *extended transition function* $\widehat{\delta} : \{0,1\}^* \to Q$ by the recursion

$$\begin{aligned} \widehat{\delta}(\lambda) &= q_0, \\ \widehat{\delta}(xb) &= \delta(\widehat{\delta}(x), b). \end{aligned}$$

For $x \in \{0,1\}^*$, we define the *output* of $T$ on $x$ to be the string $T(x)$ defined by the recursion

$$\begin{aligned} T(\lambda) &= \lambda, \\ T(xb) &= T(x)\nu(\widehat{\delta}(x), b) \end{aligned}$$

for all $x \in \{0,1\}^*$ and $b \in \{0,1\}$.

A FST can trivially act as an "optimal compressor" by outputting $\lambda$ on every transition arrow, but this is a useless compressor, because the input cannot be recovered. A FST $T = (Q, \delta, \nu, q_0)$ is *information lossless (IL)* if the function $x \mapsto (T(x), \widehat{\delta}(x))$ is one-to-one; i.e., if the output and final state of $T$ on input $x$ uniquely identify $x$. An *information lossless finite-state transducer (ILFST)* is a FST that is IL. We write ILFST to denote the set of all information lossless finite-state transducers.

Let $S \in \mathbf{C}$. The *finite-state dimension* [12] and the *finite-state strong dimension* [2] of $S$ are respectively defined

$$\dim_{\mathrm{FS}}(S) = \inf_{C \in \mathrm{ILFST}} \liminf_{n \to \infty} \frac{|C(S \upharpoonright n)|}{n},$$

and

$$\mathrm{Dim}_{\mathrm{FS}}(S) = \inf_{C \in \mathrm{ILFST}} \limsup_{n \to \infty} \frac{|C(S \upharpoonright n)|}{n}.$$

Intuitively, the finite-state dimension (resp. strong dimension) of a sequence represents the optimal best-case (resp. worst-case) compression ratio achievable on the sequence with any ILFST. We say $S \in \mathbf{C}$ is *FS-regular* if $\dim_{\mathrm{FS}}(S) = \mathrm{Dim}_{\mathrm{FS}}(S)$.

Given a sequence $P \in \mathbf{C}$ and a FST $T$, define $T(P)$ to be the *output* of $T$ on $P$, the shortest element $S \in \mathbf{C} \cup \{0,1\}^*$ such that, for all $n \in \mathbb{N}$, $T(P \upharpoonright n) \sqsubseteq S$. Let $S, P \in \mathbf{C}$ and $T \in \mathrm{ILFST}$. We say *$S$ is IL-finite-state reducible to $P$ via $T$*, and we write $S \leq_{\mathrm{ILFS}} P$ *via* $T$, if $T(P) = S$. We say *$S$ is IL-finite-state reducible to $P$*, and we write $S \leq_{\mathrm{ILFS}} P$, if there exists $T \in \mathrm{ILFST}$ such that $S \leq_{\mathrm{ILFS}} P$ via $T$. We say $S$ is *IL-finite-state equivalent* to $P$, and we write $S \equiv_{\mathrm{ILFS}} P$, if $S \leq_{\mathrm{ILFS}} P$ and $P \leq_{\mathrm{ILFS}} S$.

The following well-known theorem [26,27] shows that each ILFST $T$ computes a function $T : \mathbf{C} \to \mathbf{C}$ whose inverse is computable by another ILFST.

**Theorem 4.11** ( [26, 27]). *For all* $T \in \mathrm{ILFST}$*, there exists* $T^{-1} \in \mathrm{ILFST}$ *such that, for all* $S \in \mathbf{C}$*,* $T^{-1}(T(S)) = S$*.*

**Corollary 4.12.** *For all* $S, P \in \mathbf{C}$*,* $S \leq_{\mathrm{ILFS}} P$ *if and only if* $S \equiv_{\mathrm{ILFS}} P$*.*

The following lemma, which is similar to Observation 3.11, follows easily from Lemma 5.6 of [9]. It shows that a *single* ILFST can always be found whose compression ratio simultaneously approximates *both* of the optimal compression ratios that $\mathrm{dim}_{\mathrm{FS}}$ and $\mathrm{Dim}_{\mathrm{FS}}$ represent.

**Lemma 4.13** ( [9]). *For all* $S \in \mathbf{C}$ *and all* $\delta > 0$*, there exists an ILFST* $C$ *such that*

$$\liminf_{n\to\infty} \frac{|C(S \upharpoonright n)|}{n} < \mathrm{dim}_{\mathrm{FS}}(S) + \delta,$$

*and*

$$\limsup_{n\to\infty} \frac{|C(S \upharpoonright n)|}{n} < \mathrm{Dim}_{\mathrm{FS}}(S) + \delta,$$

The following theorem shows that ILFST's can extract finite-state dimension from sequences in a similar manner to Theorem 4.8.

**Theorem 4.14.** *For each* $\epsilon > 0$ *and each* $S \in \mathbf{C}$ *such that* $\mathrm{dim}_{\mathrm{FS}}(S) > 0$*, there exists* $P \in \mathbf{C}$ *such that* $P \equiv_{\mathrm{ILFS}} S$*,* $\mathrm{dim}_{\mathrm{FS}}(P) \geq \frac{\mathrm{dim}_{\mathrm{FS}}(S)}{\mathrm{Dim}_{\mathrm{FS}}(S)} - \epsilon$*, and* $\mathrm{Dim}_{\mathrm{FS}}(P) \geq 1 - \epsilon$*.*

*Proof.* Let $S$ and $\epsilon$ be as in the statement of the theorem. Let $\delta = \mathrm{dim}_{\mathrm{FS}}(S) \cdot \epsilon/3$. Choose $C \in \mathrm{ILFST}$ for $S$ and $\delta$ as in Lemma 4.13, so that $\liminf\limits_{n\to\infty} \frac{|C(S\upharpoonright n)|}{n} < \mathrm{dim}_{\mathrm{FS}}(S) + \delta$ and $\limsup\limits_{n\to\infty} \frac{|C(S\upharpoonright n)|}{n} < \mathrm{Dim}_{\mathrm{FS}}(S) + \delta$. Let $P = C(S)$, and let $n_p = |C(S \upharpoonright n)|$ for all $n \in \mathbb{N}$. Let $\alpha = \mathrm{dim}_{\mathrm{FS}}(S) \cdot 2/3$. Since $\mathrm{dim}_{\mathrm{FS}}(S) > 0$, for all but finitely many $n \in \mathbb{N}$, $n_p > \alpha n$. Then

$$\begin{aligned}
& \liminf_{n\to\infty} \frac{n_p}{n} < \mathrm{dim}_{\mathrm{FS}}(S) + \delta \\
\Longrightarrow \quad & \inf_{C' \in \mathrm{ILFST}} \liminf_{m\to\infty} \frac{|C'(S \upharpoonright m)|}{m} > \liminf_{n\to\infty} \frac{n_p}{n} - \delta \\
\Longrightarrow\ & (\forall C' \in \mathrm{ILFST}) \ \liminf_{m\to\infty} \frac{|C'(S \upharpoonright m)|}{m} > \liminf_{n\to\infty} \frac{n_p}{n} - \delta \\
\Longrightarrow\ & (\forall C' \in \mathrm{ILFST})(\exists^\infty n \in \mathbb{N}) \ \liminf_{m\to\infty} \frac{|C'(S \upharpoonright m)|}{m} > \frac{n_p}{n} - \delta \\
\Longrightarrow\ & (\forall C' \in \mathrm{ILFST})(\exists^\infty n \in \mathbb{N}) \ |C'(S \upharpoonright n)| > n_p - \delta n \\
\Longrightarrow\ & (\forall C' \in \mathrm{ILFST})(\exists^\infty n \in \mathbb{N}) \ |C'(S \upharpoonright n)| > n_p - \alpha\epsilon n. \qquad (4.25)
\end{aligned}$$

Let $C_2 \in \mathrm{ILFST}$. Let $C_3 = C_2 \circ C$ denote the *composition* of $C_2$ with $C$, the ILFST such that, for every $x \in \{0,1\}^*$, $C_3(x) = C_2(C(x))$. It is well-known [26, 27] that ILFST's can be composed in this manner. By (4.25), for infinitely many $n \in \mathbb{N}$,

$$\begin{aligned}
n_p - \alpha\epsilon n &< |C_3(S \upharpoonright n)| \\
&= |C_2(C(S \upharpoonright n))| \\
&= |C_2(P \upharpoonright n_p)|.
\end{aligned}$$

Thus, dividing both sides by $n_p$, since $n_p \geq \alpha n$ for all $n$,

$$\begin{aligned}\limsup_{n\to\infty} \frac{|C_2(P \upharpoonright n)|}{n} &\geq \limsup_{n\to\infty} \frac{|C_2(P \upharpoonright n_p)|}{n_p} \\ &\geq 1 - \limsup_{n\to\infty} \alpha\epsilon \frac{n}{n_p} \\ &\geq 1 - \epsilon.\end{aligned}$$

Since $C_2$ was arbitrary, this establishes that $\mathrm{Dim}_{\mathrm{FS}}(P) \geq 1 - \epsilon$.

We now show that $\dim_{\mathrm{FS}}(P) \geq \frac{\dim_{\mathrm{FS}}(S)}{\mathrm{Dim}_{\mathrm{FS}}(S)} - \epsilon$. In a similar derivation to that giving (4.25),

$$\begin{aligned}&\limsup_{n\to\infty} \frac{n_p}{n} < \mathrm{Dim}_{\mathrm{FS}}(S) + \delta \\ \Longrightarrow\ &\limsup_{n\to\infty} \frac{n_p}{n} < \dim_{\mathrm{FS}}(S) \frac{\mathrm{Dim}_{\mathrm{FS}}(S)}{\dim_{\mathrm{FS}}(S)} + \delta \\ \Longrightarrow\ &\inf_{C' \in \mathrm{ILFST}} \liminf_{m\to\infty} \frac{|C'(S \upharpoonright m)|}{m} > \frac{\dim_{\mathrm{FS}}(S)}{\mathrm{Dim}_{\mathrm{FS}}(S)} \left( \limsup_{n\to\infty} \frac{n_p}{n} - \delta \right) \\ \Longrightarrow\ &(\forall C' \in \mathrm{ILFST})(\forall^\infty n \in \mathbb{N})\ |C'(S \upharpoonright n)| > \frac{\dim_{\mathrm{FS}}(S)}{\mathrm{Dim}_{\mathrm{FS}}(S)} (n_p - \alpha\epsilon n). \qquad (4.26)\end{aligned}$$

By (4.26), for all but finitely many $n \in \mathbb{N}$,

$$\begin{aligned}\frac{\dim_{\mathrm{FS}}(S)}{\mathrm{Dim}_{\mathrm{FS}}(S)} (n_p - \alpha\epsilon n) &< |C_3(S \upharpoonright n)| \\ &= |C_2(C(S \upharpoonright n))| \\ &= |C_2(P \upharpoonright n_p)|.\end{aligned}$$

Since $C$ is an ILFST, there is a constant $c > 0$ such that, for all $n$, $((n+1)_p - n_p) \leq c$. Since $C_2$ is an ILFST, the same reasoning implies that there is a constant $c' \in \mathbb{N}$ such that, for all $n \in \mathbb{N}$, $|C_2(P \upharpoonright (n+1)_p)| - |C_2(P \upharpoonright n_p)| \leq c'$. Therefore

$$\begin{aligned}\liminf_{n\to\infty} \frac{|C_2(P \upharpoonright n)|}{n} &= \liminf_{n\to\infty} \frac{|C_2(P \upharpoonright n_p)|}{n_p} \\ &\geq \frac{\dim_{\mathrm{FS}}(S)}{\mathrm{Dim}_{\mathrm{FS}}(S)} \left( 1 - \liminf_{n\to\infty} \alpha\epsilon \frac{n}{n_p} \right) \\ &\geq \frac{\dim_{\mathrm{FS}}(S)}{\mathrm{Dim}_{\mathrm{FS}}(S)} - \epsilon.\end{aligned}$$

Since $C_2$ was arbitrary, this establishes that $\dim_{\mathrm{FS}}(P) \geq \frac{\dim_{\mathrm{FS}}(S)}{\mathrm{Dim}_{\mathrm{FS}}(S)} - \epsilon$.

Clearly, $P \leq_{\mathrm{ILFS}} S$ via $C$. By Corollary 4.12, $P \equiv_{\mathrm{ILFS}} S$. □

**Corollary 4.15.** *For each $\epsilon > 0$ and each FS-regular $S \in \mathbf{C}$ such that $\dim_{\mathrm{FS}}(S) > 0$, there exists $P \in \mathbf{C}$ such that $P \equiv_{\mathrm{ILFS}} S$ and $\dim_{\mathrm{FS}}(P) \geq 1 - \epsilon$.*

# 5 Every Sequence is Optimally Decompressible from a Random One

Kučera [28, 29] and Gács [18] independently obtained the surprising result that *every* sequence is Turing reducible to a Martin-Löf random sequence. In the words of Gács, "it permits us to view even very pathological sequences as the result of the combination of two relatively well-understood processes: the completely chaotic outcome of coin-tossing, and a transducer algorithm." Merkle and Mihailović [37] have provided a simpler proof of this result using martingales, which are strategies for gambling on successive bits of a sequence.

Bennett [4] claims that "This is the infinite analog of the far more obvious fact that every finite string is computable from an algorithmically random string (e.g., its minimal program)." However, the analogy is incomplete. Not only is every string $s$ computable from a random string $r$, but $r$ is an *optimally compact representation* of $s$. Gács showed that his reduction achieves a decompression ratio of 1: for any $n$, $n + o(n)$ bits of $R$ are required to compute $n$ bits of $S$. But as in the case of strings, sequences that are sparse in information content should in principle be derivable from a more compact description. Consider the following example. It is well known that $K$, the characteristic sequence of the halting language, has constructive dimension and constructive strong dimension 0 [3]. The binary representation of Chaitin's halting probability $\Omega = \sum_{M \text{ halts}} 2^{-|M|}$ (where $M$ ranges over all halting programs and $|M|$ is $M$'s binary description length) is an algorithmically random sequence [11]. It is known that $K \leq_{\mathrm{T}} \Omega$ (see [30]). Furthermore, only the first $n$ bits of $\Omega$ are required to compute the first $2^n$ bits of $K$, so the asymptotic decompression ratio of this reduction is 0. $\Omega$ can be considered an optimally compressed representation of $K$, and it is no coincidence that the decompression ratio of 0 achieved by the reduction is precisely the constructive dimension of $K$.

We generalize this phenomenon to arbitrary sequences, extending the result of Kučera and Gács by pushing the decompression ratio of the reduction down to its optimal lower bound. Thus, we complete Bennett's above-mentioned analogy between reductions to random sequences and reductions to random strings. We show that, for every sequence $S$, there is a sequence $R$ such that $S \leq_{\mathrm{T}} R$, where the best-case decompression ratio of the reduction is the constructive dimension of $S$, and the worst-case decompression ratio is the constructive strong dimension of $S$. Furthermore, we show that the sequence $R$ can be chosen to be Martin-Löf random, although the randomness of $R$ is easily obtained by invoking the construction of Gács in a black-box fashion. Finally, a single machine works in all cases; in analogy to the universal Turing machine used to define Kolmogorov complexity, a single Turing reduction reproduces each sequence $S$ from its shortest description. This result gives a way to associate with each sequence $S$ another sequence $R$ that is an optimally compressed representation of $S$, decompressible by a single, universal reduction machine. As in the case of Kolmogorov complexity, the compression direction is in general uncomputable; it is not always the case that $R \leq_{\mathrm{T}} S$. This result also extends a compression result of Ryabko [42, 43], discussed in section 2, although it is not a strict im-provement, since Ryabko considered two-way reductions (Turing equivalence) rather than one-way reductions.

The following lemma shows two senses in which the composition of two oracle Turing machines in a transitive Turing reduction bounds the compression ratio of the transitive reduction below the product of the compression ratios of the two original reductions.

**Lemma 5.1.** *Let $S, P, R \in \mathbf{C}$ and $M_S, M_P \in \mathrm{OTM}$ such that $S \leq_{\mathrm{T}} P$ via $M_S$ and $P \leq_{\mathrm{T}} R$ via $M_P$, and let $M = M_S \circ M_P$, so that $S \leq_{\mathrm{T}} R$ via $M$. Then*

$$\rho_M^+(S, R) \leq \rho_{M_S}^+(S, P)\rho_{M_P}^+(P, R),$$

*and*

$$\rho_M^-(S, R) \leq \rho_{M_S}^-(S, P)\rho_{M_P}^+(P, R).$$

*Proof.* Let $r_S^{P+} > \rho_{M_S}^+(S, P)$, $r_S^{P-} > \rho_{M_S}^-(S, P)$, and $r_P^{R+} > \rho_{M_P}^+(P, R)$. It suffices to show that $\rho_M^+(S, R) \leq r_S^{P+} r_P^{R+}$ and $\rho_M^-(S, R) \leq r_S^{P-} r_P^{R+}$.

For infinitely many $n$, $\#(S \upharpoonright n, M_S^P) < r_S^{P-} n$. For all but finitely many $n$, $\#(S \upharpoonright n, M_S^P) < r_S^{P+} n$, and $\#(P \upharpoonright n, M_P^R) < r_P^{R+} n$. Then, for all but finitely many $n$, to compute $S \upharpoonright n$, $M^R$ requires

$$\#(S \upharpoonright n, M^R) = \#\left(P \upharpoonright \#\left(S \upharpoonright n, M_S^P\right), M_P^R\right) < r_P^{R+} \#\left(S \upharpoonright n, M_S^P\right) < r_S^{P+} r_P^{R+} n$$

queries to $R$. Since this holds for all but finitely many $n$,

$$\rho_M^+(S, R) = \limsup_{n \to \infty} \frac{\#(S \upharpoonright n, M^R)}{n} \leq r_S^{P+} r_P^{R+}.$$

For infinitely many $n$, to compute $S \upharpoonright n$, $M$ requires

$$\#(S \upharpoonright n, M^R) = \#\left(P \upharpoonright \#\left(S \upharpoonright n, M_S^P\right), M_P^R\right) < r_P^{R+} \#\left(S \upharpoonright n, M_S^P\right) < r_S^{P-} r_P^{R+} n$$

queries to $R$. Since this holds for infinitely many $n$,

$$\rho_M^-(S, R) = \liminf_{n \to \infty} \frac{\#(S \upharpoonright n, M^R)}{n} \leq r_S^{P-} r_P^{R+}.$$

□

The following may be proven using the same technique as the proof of Lemma 4.4.

**Lemma 5.2.** *Let $S, R \in \mathbf{C}$ and $M \in \mathrm{OTM}$ such that $S \leq_{\mathrm{T}} R$ via $M$. Then*

$$\rho_M^-(S, R) \geq \dim(S), \text{ and } \rho_M^+(S, R) \geq \mathrm{Dim}(S).$$

The next lemma is similar to Lemma 4.6, but it achieves a more compact encoding of the rational number at the expense of a worse approximation to the real it is representing. This extra compactness is required in order to make the length of the program for the rational number asymptotically smaller than the length of the block in which it resides, so that its length can be entirely discounted in the limit (not just made an arbitrarily small percentage of the block length, as in Lemma 4.6).

**Lemma 5.3.** *There exists a constant $C \in \mathbb{N}$ such that the following holds. For all $i \in \mathbb{N}$, let $r_i \in [1, 2^{i^2}]$. Then for all but finitely many $i \in \mathbb{N}$, there is a rational number $c_i \in \mathbb{Q}^+$ such that $r_i\left(1 - \frac{1}{i^2}\right) \leq c_i < r_i$ and $\mathrm{K}(c_i) \leq C \log i$.*

*Proof.* We prove the cases $r_i \geq i^2$ and $1 \leq r_i < i^2$ separately. Suppose $r_i \geq i^2$. In this case we will choose $c_i$ to be an integer. Set $m \in \mathbb{Z}^+$ such that $2^{m-1} < i^2 \leq 2^m$. Since $r_i \geq i^2 > 2^{m-1}$, $\lceil \log r_i \rceil > m - 1$.

Let $c_i \in \mathbb{Z}^+$ be the integer whose binary representation is $x0^{\lceil \log r_i \rceil - m}$, where $x \in \{0,1\}^m$ is the first $m$ bits of $\lfloor r_i \rfloor$. Since $c_i$ shares its first $m$ bits with $r_i$,

$$r_i - c_i \leq 2^{\lceil \log r_i \rceil - m} - 1 \leq \frac{r_i + 2}{2^m} - 1 \leq \frac{r_i}{i^2},$$

so $r_i\left(1 - \frac{1}{i^2}\right) \leq c_i < r_i$. $c_i$ can be fully described by the first $m$ bits of $r_i$, along with the binary representation of the number $\lceil \log r_i \rceil - m$ of 0's that follow. Thus, describing $c_i$ requires no more than $m + \log(\lceil \log r_i \rceil - m) \leq \log i^2 + 1 + \log\log c + \log i^2 = O(\log i)$ bits.

This will not work if $r_i \in \mathbb{Z}^+$ and $r_i$'s least significant $\lceil \log r_i \rceil - m$ bits are 0, which would result in $c_i = r_i$, rather than $c_i < r_i$. In this case, let

$$c_i = r_i - 1 = \mathrm{num}_2\left(\mathrm{rep}_2(\mathrm{num}_2(x) - 1)1^{\lceil \log r_i \rceil - m}\right),$$

where $\mathrm{num}_2(x)$ is the integer whose binary representation is $x$, and $\mathrm{rep}_2(n)$ is the binary representation (with possible leading zeroes) of $n \in \mathbb{N}$. This likewise requires $O(\log i)$ bits to describe. Since $r_i \geq i^2$, $c_i = r_i - 1 \geq r_i\left(1 - \frac{1}{i^2}\right)$.

Now suppose that $1 \leq r_i < i^2$. We approximate $r_i$ by the binary integer $\lfloor r_i \rfloor$, plus a finite prefix of the bits to the right of $r_i$'s decimal point in binary form. If $x.S$ is the binary representation of $r_i$, where $x \in \{0,1\}^*$ and $S \in \mathbf{C}$, let $c_i \in \mathbb{Z}^+$ be represented by $x.y$, where $y \sqsubset S$.

Since $r_i < i^2$, $|x| \leq \log i^2 = O(\log i)$. We need $r_i - c_i \leq \frac{r_i}{i^2}$ for $c_i$ to approximate $r_i$ closely. Since $r_i - c_i \leq 2^{-|y|}$, it suffices to choose $y \sqsubset S$ such that $2^{-|y|} \leq \frac{r_i}{i^2}$, or $|y| \geq \log \frac{i^2}{r_i}$. Let $|y| = \left\lceil \log \frac{i^2}{r_i} \right\rceil = O(\log i)$, since $r_i \geq 1$. Thus $|x| + |y| = O(\log i)$, so describing $c_i$ requires $O(\log i)$ bits.

This will not work if $r_i$ is a dyadic rational $x.z$, where $x, z \in \{0,1\}^*$ and $|z| \leq |y|$, which would result in $c_i = r_i$, rather than $c_i < r_i$. In this case, let $r_i' \in \left[r_i\left(1 - \frac{1}{2i^2}\right), r_i\right)$ be irrational. Choose $c_i$ for $r_i'$ by the method just described, such that $r_i' > c_i \geq r_i'\left(1 - \frac{1}{2i^2}\right)$, and $c_i$ requires $O(\log(i\sqrt{2})) = O(\log i)$ bits. Then $c_i \geq r_i\left(1 - \frac{1}{i^2}\right)$ by the triangle inequality, and $c_i < r_i' < r_i$. □

The next theorem says that every sequence is Turing reducible to a random sequence. Part 1 is due independently to Kučera and Gács, and part 2 is due to Gács.

**Theorem 5.4** ( [18, 28, 29]). *There is an OTM $M_{\mathrm{g}}$ such that, for all $S \in \mathbf{C}$, there is a sequence $R \in \mathsf{RAND}$ such that*

1. $S \leq_{\mathrm{T}} R$ *via* $M_{\mathrm{g}}$.

2. $\rho^+_{M_{\mathrm{g}}}(S, R) = 1$.

The next theorem is the main result of this section. It shows that the compression lower bounds of Lemma 5.2 are achievable, and that a single OTM $M$ suffices to carry out the reduction, no matter which sequence $S$ is being computed. Furthermore, the oracle sequence $R$ to which $S$ reduces can be made Martin-Löf random. The randomness of $R$ is easily accomplished by invoking the construction of Gács in a black-box fashion; the majority of the work in the proof is establishing the bound on the compression. Therefore, the next theorem improves the query usage bound of Theorem 5.4, but it does not provide a new proof of Theorem 5.4. The proof is similar to the proof of Theorem 4.5. However, due to the existence of the optimal constructive martingale $\mathbf{d}$ and the fact that we require only the decompression direction to be computable, we need not approximate the optimal decompression ratio from above; we can actually achieve it using $\mathbf{d}$.

**Theorem 5.5.** *There is an OTM $M$ such that, for all $S \in \mathbf{C}$, there is a sequence $R \in$ $\mathsf{RAND}$ such that*

1. $S \leq_{\mathrm{T}} R$ *via* $M$.

2. $\rho^-_M(S, R) = \dim(S)$.

3. $\rho^+_M(S, R) = \mathrm{Dim}(S)$.

*Proof.* Parts 1 and 3 are shown. The proof for part 2 is similar to part 3.

If $S \in \mathsf{RAND}$, then $S \leq_{\mathrm{T}} S$ via the trivial "bit copier" machine $M'$, with lower and upper compression ratio $\dim(S) = \mathrm{Dim}(S) = 1$, so assume that $S \not\in \mathsf{RAND}$.

A single OTM $M''$ suffices to carry out the reduction described below, no matter what sequence $S \not\in \mathsf{RAND}$ is being computed. If $S \in \mathsf{RAND}$, then $M'$ is used. These two separate reductions are easily combined into one by reducing each sequence $S$ to a random sequence $bR$ via $M \in$ OTM, where $b \in \{0, 1\}$, $R = S$ if $S \in \mathsf{RAND}$, and $R$ is given by the construction below if $S \not\in \mathsf{RAND}$. The bit $b$ indicates to $M$ whether to use $M'$ or $M''$ for the reduction. Hence a single OTM $M$ implements the "optimal decompression".

For all $i \in \mathbb{N}$, define $k_i = i + 1$, and define $n_0 = 1$ and $n_i = n_{i-1} + k_i = \frac{(i+1)(i+2)}{2}$ for $i > 0$. Note that $n_i \leq i^2$ for all sufficiently large $i$. Write $S = \widehat{s}_0\widehat{s}_1 \ldots$, such that, for all $i \in \mathbb{N}$, $|\widehat{s}_i| = k_i$; $k_i$ represents the length of the $i^{\mathrm{th}}$ block into which we subdivide $S$. $n_i$ is then $|\widehat{s}_0 \ldots \widehat{s}_i|$, the total length of the first $i + 1$ blocks. For all $i \in \mathbb{N}$, define $c_i \in \mathbb{Q}^+$ to be a rational number satisfying

1. $\mathbf{d}(S \restriction n_i)\left(1 - \frac{1}{i^2}\right) \leq c_i < \mathbf{d}(S \restriction n_i)$; i.e., $c_i$ is a rational number approximating $\mathbf{d}(S \restriction n_i)$ from below.

2. $\mathrm{K}(c_i) = o(k_i)$ as $i \to \infty$; i.e. $c_i$ can be computed from a program asymptotically smaller than the length of the $i^{\text{th}}$ block.

By Observation 3.13, $1 \leq \mathbf{d}(\lambda) < 2$. By Observation 3.1, $\mathbf{d}(S \upharpoonright n_i) \leq 2^{n_i}\mathbf{d}(\lambda) \leq 2^{i^2}$ for sufficiently large $i$. By Observation 3.6, $S \notin \mathsf{RAND}$ implies that for all but finitely many $i$, $\mathbf{d}(S \upharpoonright n_i) \geq 1$. Thus, by Lemma 5.3 (take $r_i = \mathbf{d}(S \upharpoonright n_i)$), there is a $c_i \in \mathbb{Q}^+$ satisfying the above two conditions.

For all $i \in \mathbb{N}$, define the set $A_i = A^{(k_i)}_{\mathbf{d},c_i,S\upharpoonright n_{i-1}} \subseteq \{0,1\}^{k_i}$ as in Lemma 4.7 by

$$A_i = \left\{ u \in \{0,1\}^{k_i} \;\middle|\; \mathbf{d}((S \upharpoonright n_{i-1})u) > c_i \right\},$$

the set of all length-$k_i$ extensions of $S \upharpoonright n_{i-1}$ that add more capital to the optimal constructive martingale $\mathbf{d}$ than $\widehat{s}_i i - 1]$ does, using $c_i$ as an approximation to $\mathbf{d}(S \upharpoonright n_i)$. Since $\mathbf{d}(S \upharpoonright n_i) > c_i$, it follows that $\widehat{s}_i \in A_i$.

For all $i \in \mathbb{N}$, let $p_i \in \mathbb{N}$ be the output of the following partial computable procedure, when given as input the string $\widehat{s}_i - 1] \in \{0,1\}^{k_i}$:

$\mathbf{ind}^{(k_i)}_{c_i,S\upharpoonright n_{i-1}}(\widehat{s}_i - 1] \in \{0,1\}^{k_i})$

$A_i \leftarrow \varnothing$
**for** $t = 0, 1, 2, \ldots$
  **do for** each $u \in \{0,1\}^{k_i} - A_i$
    **do if** $\widehat{\mathbf{d}}((S \upharpoonright n_{i-1})u, t) > c_i$
      **then** add $u$ to $A_i$
        **if** $u = \widehat{s}_i$
          **then** output $|A_i|$ and halt

In other words, $p_i$ is the order in which $\mathbf{d}(S \upharpoonright n_i)$ is shown to exceed $c_i$ (i.e., to belong to $A_i$) by a parallel evaluation of $\widehat{\mathbf{d}}((S \upharpoonright n_{i-1})u, t)$ on all extensions $u \in \{0,1\}^{k_i}$ of $S \upharpoonright n_{i-1}$, for $t = 0, 1, 2, \ldots$. Since $c_i < \mathbf{d}(S \upharpoonright n_i)$, there exists some $t \in \mathbb{N}$ such that $\widehat{\mathbf{d}}(S \upharpoonright n_i, t) > c_i$, and so $p_i$ is well-defined. The computation of $\mathbf{str}^{(k_i)}_{c_i,S\upharpoonright n_{i-1}}$, the inverse of $\mathbf{ind}^{(k_i)}_{c_i,S\upharpoonright n_{i-1}}$, resembles that of $\mathbf{ind}^{(k_i)}_{c_i,S\upharpoonright n_{i-1}}$:

$\mathbf{str}^{(k_i)}_{c_i,S\upharpoonright n_{i-1}}(p_i \in \mathbb{N})$

$A_i \leftarrow \varnothing$
**for** $t = 0, 1, 2, \ldots$
  **do for** each $u \in \{0,1\}^{k_i} - A_i$
    **do if** $\widehat{\mathbf{d}}((S \upharpoonright n_{i-1})u, t) > c_i$
      **then** add $u$ to $A_i$
        **if** $|A_i| = p_i$
          **then** output $u$ and halt

Note that $\mathbf{str}^{(k_i)}_{c_i,S\upharpoonright n_{i-1}}$ will not halt if given as input an integer greater than $|A_i|$, and $\mathbf{ind}^{(k_i)}_{c_i,S\upharpoonright n_{i-1}}$ will not halt if given a string that is not an element of $A_i$.

$\mathbf{ind}^{(k_i)}_{c_i,S\upharpoonright n_{i-1}}$ and $\mathbf{str}^{(k_i)}_{c_i,S\upharpoonright n_{i-1}}$ are similar to $\mathrm{ind}^{(k)}_{d,c,s}$ and $\mathrm{str}^{(k)}_{d,c,s}$, respectively, but are designed explicitly to work with the lower-semicomputable martingale $\mathbf{d}$. Instead of enumerating strings in lexicographic order, they use the fact that the lower graph of $\mathbf{d}$ is computably enumerable via $\widehat{\mathbf{d}}$ to enumerate strings in the set $A_i$.

For all $i \in \mathbb{N}$, let $\pi(c_i)$ denote a self-delimiting, shortest program for computing $c_i$. Define the sequence $P \in \mathbf{C}$ by

$$P = \mathrm{enc}(p_0)\pi(c_0)\mathrm{enc}(p_1)\pi(c_1)\mathrm{enc}(p_2)\pi(c_2)\ldots.$$

Define the oracle Turing machine $M_S$ that produces $n$ bits of $S$, with oracle $P$, as follows. Let $i(n)$ denote the block in which $n$ resides – the unique $i \in \mathbb{N}$ such that $n_i \leq n < n_{i+1}$. First, $M_S^P$ reads the first $i(n)+1$ blocks of $P$:

$$\mathrm{enc}(p_0)\pi(c_0)\ldots\mathrm{enc}(p_{i(n)})\pi(c_{i(n)}).$$

$M_S^P$ then calculates the first $i(n)+1$ blocks of $S$ iteratively. On block $i$, $M_S^P$ first computes $p_i$ from $\mathrm{enc}(p_i)$ and $c_i$ from $\pi(c_i)$. Then, $M_S^P$ evaluates $\mathbf{str}^{(k_i)}_{c_i,S\upharpoonright n_{i-1}}(p_i)$ to obtain $\widehat{s}_i$ and outputs it as the $i^{\text{th}}$ block of $S$.

Since $\widehat{s}_i \in A_i$, it follows that $p_i \leq |A_i|$, and so $|\mathrm{enc}(p_i)| \leq \log|A_i| + 2\log\log|A_i| + 3$. Note that

$$\begin{aligned}
-\sum_{j=2}^{i}\log\left(1-\frac{1}{j^2}\right) &= -\sum_{j=2}^{i}\log\frac{(j+1)(j-1)}{j^2} \\
&= -\sum_{j=2}^{i}\left(\log(j+1)+\log(j-1)-2\log j\right) \\
&= -\log 1 + \log 2 + \log i - \log(i+1),
\end{aligned}$$

which converges to 1 as $i \to \infty$, so $g(j) = 1 - \frac{1}{j^2}$ satisfies $-\sum_{j=2}^{i}\log g(j) = o(n_i)$, whence the conditions of Lemma 4.7 are satisfied. Therefore, by Lemma 4.7,

$$\limsup_{i\to\infty}\frac{\sum_{j=0}^{i}|\mathrm{enc}(p_j)|}{n_i} \leq \limsup_{i\to\infty}\frac{\sum_{j=0}^{i}\log|A_j|}{n_i} \leq \mathrm{Dim}(S). \tag{5.1}$$

By our choice of $c_i$, $|\pi(c_i)| = o(k_i)$, so $\sum_{j=0}^{i}|\pi(c_j)| = o(n_i)$ as $i \to \infty$, giving

$$\limsup_{i\to\infty}\frac{\sum_{j=0}^{i}|\mathrm{enc}(p_j)\pi(c_j)|}{n_i} = \limsup_{i\to\infty}\frac{\sum_{j=0}^{i}|\mathrm{enc}(p_j)|}{n_i}. \tag{5.2}$$

Since $n_i = \frac{k_i(k_i+1)}{2}$, $k_i = o(n_i)$, so

$$\limsup_{n\to\infty}\frac{\sum_{j=0}^{i(n)}|\mathrm{enc}(p_j)\pi(c_j)|}{n} \leq \limsup_{i\to\infty}\frac{\sum_{j=0}^{i}|\mathrm{enc}(p_j)\pi(c_j)|}{n_i}. \tag{5.3}$$

In other words, because the block size grows slower than the prefix length, the lim sup over all blocks is at least the lim sup over all bits (and they are in fact equal by the definition of lim sup).

For all $n \in \mathbb{N}$, $M_S^P$ requires $\sum_{j=0}^{i(n)} |\mathrm{enc}(p_j)\pi(c_j)|$ bits of $P$ in order to compute $n$ bits of $S$, and hence, by (5.1)-(5.3),

$$\rho^+_{M_S}(S, P) = \limsup_{n\to\infty} \frac{\sum_{j=0}^{i(n)} |\mathrm{enc}(p_j)\pi(c_j)|}{n} \leq \mathrm{Dim}(S).$$

Choose $R \in \mathsf{RAND}$ for $P$ as in the construction of Gács in his proof of Theorem 5.4, satisfying $P \leq_{\mathrm{T}} R$ via $M_{\mathrm{g}}$ and $\rho^+_{M_{\mathrm{g}}}(P, R) = 1$. Let $M'' = M_S \circ M_{\mathrm{g}}$. Then $S \leq_{\mathrm{T}} R$ via $M''$ and, by Lemma 5.1,

$$\rho^+_{M''}(S, R) \leq \rho^+_{M_S}(S, P)\rho^+_{M_{\mathrm{g}}}(P, R) \leq \mathrm{Dim}(S).$$

By Lemma 5.2, $\rho^+_{M''}(S, R) \geq \mathrm{Dim}(S)$. Similarly, $\rho^-_{M''}(S, R) = \dim(S)$. □

Note that the block lengths used in the proof grow with the square root of the prefix length; if we write $S = \widehat{s}_0\widehat{s}_1\ldots$ as in the proof of Theorem 5.5, then for all $i \in \mathbb{N}$, $|\widehat{s}_i| = O(\sqrt{|\widehat{s}_0 \ldots \widehat{s}_i|})$, and similarly for the blocks of $R$. Bienvenu [6] has shown that this can be improved to $|\widehat{s}_i| = O(\log |\widehat{s}_0 \ldots \widehat{s}_i|)$.

It is instructive to compare Theorem 5.5 with Ryabko's Theorem 4.1. While Ryabko's theorem represents $S$ with a more compact sequence $R$, it is not optimally compact, as a different decoding machine is required to get the compression ratio closer and closer to the optimal ratio of $\dim(S)$. However, the major difference between the theorems is that Ryabko's construction does not achieve the bound between $\rho^+$ and Dim. Intuitively, Ryabko's theorem states that $S$ may be compressed to a sequence $R$, where *infinitely often* (but not almost everywhere), approximately the first $\mathrm{K}(S \upharpoonright n)$ bits of $R$ suffice to produce $S \upharpoonright n$. However, Ryabko's construction requires that the block lengths grow exponentially: for all $i \in \mathbb{N}$, $|\widehat{s}_0 \ldots \widehat{s}_{i+1}| > 2^i|\widehat{s}_0 \ldots \widehat{s}_i|$. Therefore, while the lower compression ratio $\rho^-$ is close to optimal in Ryabko's theorem, the upper compression ratio $\rho^+$ is infinite.

# 6 Dimension Characterizations

We use the decompression results of previous sections to characterize constructive dimension as the optimal decompression ratio achievable on a sequence with Turing reductions, to characterize computable dimension as the optimal decompression ratio achievable on the sequence with truth-table reductions, and to characterize $\mathrm{p}_i$space-dimension as the optimal decompression ratio achievable on the sequence with $\mathrm{p}_i$space-computable Turing reductions. See [14] for a collection of similar compression and decompression characterizations of finite-state dimension.

**Theorem 6.1.** *For all $i \in \mathbb{N}$ and $S \in \mathbf{C}$,*

$$\begin{array}{lcl} \dim(S) & = & \rho^{-}(S), \\ \mathrm{Dim}(S) & = & \rho^{+}(S), \\ \dim_{\mathrm{comp}}(S) & = & \rho^{-}_{\mathrm{tt}}(S), \\ \mathrm{Dim}_{\mathrm{comp}}(S) & = & \rho^{+}_{\mathrm{tt}}(S), \\ \dim_{\mathrm{p}_i\mathrm{space}}(S) & = & \rho^{-}_{\mathrm{p}_i\mathrm{space}}(S), \\ \mathrm{Dim}_{\mathrm{p}_i\mathrm{space}}(S) & = & \rho^{+}_{\mathrm{p}_i\mathrm{space}}(S). \end{array}$$

*Proof.* This follows immediately from Theorems 4.5 and 5.5 and Lemmas 4.4 and 5.2. □

Note that $\mathrm{p}_i$-dimension is absent from Theorem 6.1, because Lemma 4.4 is not known to hold for $\mathrm{p}_i$-dimension without a Kolmogorov complexity characterization of $\mathrm{p}_i$-dimension. Theorem 3.14, however, characterizes $\mathrm{p}_i$-dimension in terms of optimal $\mathrm{p}_i$-computable decompression if $\mathrm{p}_i$-computable compression (i.e., reversibility) is also required.

These decompression characterizations differ from Mayordomo's and Hitchcock's Kolmogorov complexity characterizations of these dimensions in that the compressed version of a prefix of $S$ does not change drastically from one prefix to the next, as it would in the case of Kolmogorov complexity. While the theory of Kolmogorov complexity assigns to each finite string an optimally compact representation of that string – its shortest program – this does not easily allow us to compactly represent an infinite sequence with another infinite sequence. This contrasts, for example, the notions of finite-state compression [26] or Lempel-Ziv compression [54], which are *monotonic*: for all strings $x$ and $y$, $x \sqsubseteq y$ implies that $C(x) \sqsubseteq C(y)$, where $C(x)$ is the compressed version of $x$. Monotonicity enables these compression algorithms to encode and decode an infinite sequence – or in a more applied setting, a data stream of unknown length – online, without needing to reach the end of the data before starting. However, if we let $\pi_x$ and $\pi_y$ respectively be shortest programs for $x$ and $y$, then $x \sqsubseteq y$ does not imply that $\pi_x \sqsubseteq \pi_y$.[3] In fact, it may be the case that $\pi_x$ is longer than $\pi_y$, or that $\pi_x$ and $\pi_y$ do not even share any prefixes in common. In the self-delimiting formulation of Kolmogorov complexity, $\pi_x$ *cannot* be a prefix of $\pi_y$.

The Turing reductions of Theorems 5.4, 4.1, and 5.5 satisfy the stronger properties of the *weak truth-table reduction* (see [48]), which is a Turing reduction in which the query usage of the OTM on input $n$ is bounded by a computable function of $n$. Thus, constructive dimension and strong dimension could also be defined in terms of decompression via weak truth-table reductions (i.e., $\dim(S) = \rho^{-}_{\mathrm{wtt}}(S)$ and $\mathrm{Dim}(S) = \rho^{+}_{\mathrm{wtt}}(S)$).

**Acknowledgments.** I thank Philippe Moser, Xiaoyang Gu, Pavan Aduri, Satyadev Nandakumar, Scott Summers, Frank Stephan, and Laurent Bienvenu for helpful discussions, and Fengming Wang for pointing out [52]. I also thank anonymous referees for their helpful suggestions, Jack Lutz and Jim Lathrop for their helpful advice in preparing this article, and John Hitchcock for making useful corrections in an earlier draft, and for helpful discussions.

[3]Monotone complexity [45, 55] does not resolve this, since the universal monotone machine guarantees only the converse implication: $\pi_x \sqsubseteq \pi_y \implies x \sqsubseteq y$.